%% file: main-SIGGRAPH.tex
\documentclass[sigconf]{acmart}

\usepackage{caption}
\usepackage{subcaption}
\usepackage{multirow}
\usepackage{multicol}

\AtBeginDocument{%
  }
\setcopyright{acmlicensed}
\copyrightyear{2025}
\acmConference[SIGGRAPH Conference Papers ’25]{}{August 10–14, 2025}{Vancouver, BC, Canada}

\acmSubmissionID{718}

\begin{document}

\title{Multi-Person Interaction Generation from Two-Person Motion Priors}


\author{Wenning Xu}
\email{2878278x@student.gla.ac.uk}
\affiliation{%
  \institution{University of Glasgow}
  \city{Glasgow}
  \country{United Kingdom}
}

\author{Shiyu Fan}
\email{s.fan.1@research.gla.ac.uk}
\affiliation{%
  \institution{University of Glasgow}
  \city{Glasgow}
  \country{United Kingdom}
}

\author{Paul Henderson}
\email{paul@pmh47.net}
\affiliation{%
  \institution{University of Glasgow}
  \city{Glasgow}
  \country{United Kingdom}
}

\author{Edmond S. L. Ho}
\authornote{Corresponding author}
\email{Shu-Lim.Ho@glasgow.ac.uk}
\affiliation{%
  \institution{University of Glasgow}
  \city{Glasgow}
  \country{United Kingdom}
}


\input{sec/0_abstract}


\begin{teaserfigure}
  \includegraphics[width=\textwidth]{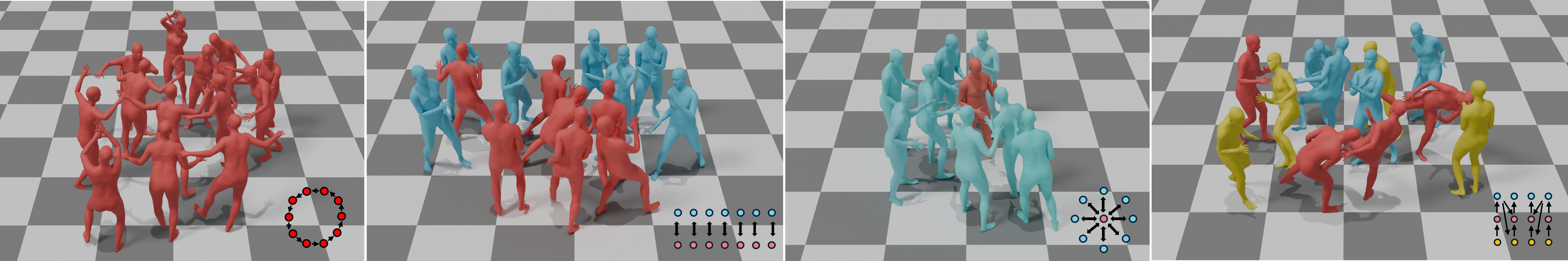}
  \caption{Examples of multi-person close interactions generated using the proposed Graph-driven Interaction Sampling. Different graph topologies were used to represent different kinds of interaction patterns. Note the Pairwise Interaction Graph illustrated at the bottom-right in each example, which lets the user control which characters are taking part in close pairwise interactions. 
  } 
  \Description{tbc.}
  \label{fig:teaser}
\end{teaserfigure}


\maketitle

\input{sec/1_intro}
\input{sec/2_related_work}
\input{sec/3_Preliminaries}
\input{sec/4_method}

\input{sec/5_experiments}
\input{sec/6_discussion}

\bibliographystyle{ACM-Reference-Format}
\bibliography{main}

\clearpage
\input{sec/figure-only}

\clearpage
\section*{Supplementary Material}
\appendix
\input{sec/X_suppl}

\end{document}

%% file: sec/0_abstract.tex
\begin{abstract}
Generating realistic human motion with high-level controls is a crucial task for social understanding, robotics, and animation. With high-quality MOCAP data becoming more available recently, a wide range of data-driven approaches have been presented. However, modelling multi-person interactions still remains a less explored area.
In this paper, we present Graph-driven Interaction Sampling, a method that can generate realistic and diverse multi-person interactions by leveraging existing two-person motion diffusion models as motion priors. Instead of training a new model specific to multi-person interaction synthesis, our key insight is to spatially and temporally separate complex multi-person interactions into a graph structure of two-person interactions, which we name the Pairwise Interaction Graph. We thus decompose the generation task into simultaneous single-person motion generation conditioned on one other's motion. 
In addition, to reduce artifacts such as interpenetrations of body parts in generated multi-person interactions, we introduce two graph-dependent guidance terms into the diffusion sampling scheme. Unlike previous work, our method can produce various high-quality multi-person interactions without having repetitive individual motions. Extensive experiments demonstrate that our approach consistently outperforms existing methods in reducing artifacts when generating a wide range of two-person and multi-person interactions.

\end{abstract}

%% file: sec/1_intro.tex
\section{Introduction}

In recent years, text-guided human motion generation has made rapid advancements, becoming a popular area in generative modelling. By generating natural human motions from text prompts, this task holds significant value across applications such as animation, gaming, and virtual reality. Especially with progress in diffusion models, text-to-motion models can more accurately translate complex language descriptions into dynamic motion sequences \cite{tevet2022human, ji2024text, wang2024move, liang2024intergen, yi2025generating, sun2024lgtm}. Despite notable progress, most existing works focus on single-person settings, with limited exploration of the complex and rich interactions that occur between multiple people.

Previous interaction generation approaches mostly consider two-person interaction generation. ComMDM \cite{shafir2023human} fine-tune the human motion diffusion model (MDM) \cite{tevet2022human} for dual human generation but they suffer from poor and limited datasets. \citet{liang2024intergen} capture a large two-person interaction dataset, and propose a parallel shared-weight diffusion model to generate two-person interactions.  While high-quality two-person interactions can now be generated by these methods trained using recently published interaction datasets such as Inter-X \cite{xu2024inter}, there are very few approaches proposed for generating multi-person interactions \cite{fan2024freemotion, shan2024towards}. 

Unlike large-scale motion capture datasets that record humans in isolation or pairs, datasets with multi-person interactions are limited, which makes it difficult to directly extend the current single- or two-person motion synthesis pipelines to support multi-person scenarios. Moreover, current approaches are mostly designed for a fixed number of people in the interaction. When the number of involved individuals has to be changed, a new model using the specific dataset with that number of people \cite{SocialDiffusion:ICCV2023} needs to be retrained. On the other hand, we believe two-person motions provide valuable motion priors for multi-person interactions. A recent work, FreeMotion \cite{fan2024freemotion}, shares a similar interest in using a two-person motion dataset for multi-person interaction generation. FreeMotion generates the motion of each person in the interaction sequentially. While such a structure theoretically allows the model to generate multi-person interactions, in practice only simple and repetitive interactions were demonstrated in the results, with the relationship between individuals being chaotic.  

We propose Graph-driven Interaction Sampling, which is capable of generating complex multi-person interactions involving different numbers of people without further training, using only existing two-person models.
Our key insight is that multi-person interactions can be modelled using a graph.
For example, in a scene where a protagonist fights off many attackers, there is a complex interaction between each attacker and the defender, while the different attackers are less strongly interacting. This situation can be described by a graph where every opponent links to the main character (Fig.~\ref{fig:teaser} second from the right). Alternatively, some social interactions can be represented using a cyclic graph to connect all characters in the scene (Fig.~\ref{fig:teaser} leftmost), while different team fights can be represented by more complex graphs (Fig.~\ref{fig:teaser} rightmost and second from the left).
Thus, we can decompose a multi-person interaction into a composition of two-person interactions between graph neighbors, with each person's motion conditioned only on that of one other---exactly the situation that two-person interaction models are already trained for. 

Decomposing multi-person motion into two-person interactions enables the use of existing models. However, we cannot simply disregard relations between persons not directly connected by a graph edge---doing so leads to intersections and other artifacts.
We mitigate these by introducing additional collision losses between non-interacting persons, and using these to guide the diffusion sampling process.
In addition, we incorporate guidance based on Gauss linking integral (GLI) \cite{Pohl:GLI} for closely-interacting persons, mitigating intersections created by the underlying two-person generation models.

Our main contributions are summarized below:
\begin{enumerate}
    \item We propose Graph-driven Interaction Sampling to generate interactions involving variable numbers of people using the existing two-person motion diffusion model. 

    \item We introduce a novel interaction sampling optimization process to effectively reduce collisions and interpenetrations in the generated 
    interactions by introducing Proxemics and Gauss Linking Integral (GLI) losses. 
    \item The results of our experiments demonstrate that we outperform current interaction generation methods. Our method generates a wide variety of highly controllable close interactions with more than ten characters.
\end{enumerate}

%% file: sec/2_related_work.tex
\section{Related Work}
\label{sec:related work}

In this section, we will first review previous research on synthesizing two-character interactions (Section~\ref{sec:2-ppl}). Next, approaches for generating interactions with multiple characters will be discussed (Section~\ref{sec:multi-ppl}). Finally, Motion Diffusion Models which are the core methodologies in the recent state-of-the-art human motion synthesis models will be presented in Section~\ref{sec:MDM}.

\subsection{Two-character Interactions} \label{sec:2-ppl}
Two decades ago, high-quality MOCAP data became more accessible, with publicly available datasets such as the CMU MOCAP dataset \cite{mocapdata}; however the majority of the data were captured from a single subject. Only very limited two-person social interactions and dancing motions were available.
This motivated a stream of early work on re-proposing singly captured MOCAP data in synthesizing two-character close interactions. \citet{Liu:SCA06:PhyInteraction} proposed using spacetime optimisation to generate physically realistic close interactions of characters by optimizing dynamics and interaction 
constraints simultaneously. In \cite{Ho:PG07}, topological constraints were used to avoid interpenetrations of body parts in wrestling motions synthesis. \citet{shum:VRST07} proposed using action-level motion graph with temporal expansion to synthesize competitive interactions such as kickboxing.  
In addition, users can also interact with virtual characters such as performing partner dance \cite{Ho:TOMM13} and sword fight \cite{Dehesa:CHI20} in VR.

Since then, researchers have focused on resolving artifacts such as loss of body contacts and interpenetrations of body parts in two-character interaction synthesis. The key idea is to minimize the changes in the spatial relationships 
among the \textit{feature points} sampled from body joints \cite{Ho:ToG10,Zhang:SIGGRAPH2023:interaction} and skin/mesh \cite{Jin:CFG18,Jang:SA2024:retarget} while editing the interactions. \citet{Zhang:SIGGRAPH2023:interaction} further utilize the spatial relation representation in the reward function to facilitate control policies learning  
in deep reinforcement learning (DRL) for retargeting characters. \citet{Won:SIGGRAPH2021:DRL} trained the imitation policy for individual characters using singly captured motions and the competitive policy for synthesizing two-character interactions using DRL.

On the other hand, learning from motion data has become the mainstream for close interaction generation in recent years as two-person interaction datasets \citep{Shen:TVCG2020,ExPI:CVPR2022,CHI3D:CVPR2020,liang2024intergen} are more accessible. In addition, synthetic data can also be used in 
interactive motion model training as demonstrated in \cite{Jang:SA2024:retarget,Li:CVPRW2024:interaction}. Social Diffusion \cite{SocialDiffusion:ICCV2023} was used for modelling the distribution of multi-person interactions using a diffusion-based framework. \citet{DuMMF:ICLR2023} proposed DuMMF to model social interactions using a dual-level model, with the local level for individual motion and the global level for social interactions.

\subsection{Multi-character Interactions} \label{sec:multi-ppl}
Compared with two-character interactions, multi-character interaction synthesis is a much less-explored area, mostly due to the lack of dataset available in the literature. UMPM Benchmark~\cite{HICV11:UMPM} is one of the earliest publicly available multi-person MOCAP datasets which contains 9 scenarios on social interactions 
recorded from 1-4 subjects. CMU Panoptic Studio \cite{Joo_2015_ICCV} provide 3D key points of 1-8 people in various social interaction scenarios. The MuPoTs-3D \cite{Mehta:3DV2018} dataset captured interactions such as social interactions, fighting and exercises from up to 3 subjects.  As a results, existing multi-character intersection synthesis approaches are still based on combining motions \cite{shum2008interaction,won2014generating, kim2012tiling, fan2024freemotion}. Patch-based methods \cite{shum2008interaction, kim2012tiling} compose multi-person interactions by spatio-temporally aligning and stitching small patches with a two-character interaction in each patch. 
\citet{won2014generating} propose a graph-based scene representation to construct multi-person interactions by single- and two-person motions. However, ours is flexible for a wide range of interaction patterns (e.g. 1-vs-1, 1-vs-many, etc) while each character can only interact with one other character at a time in \cite{won2014generating}. 

\subsection{Motion Diffusion Models} \label{sec:MDM} 
Recent progress in Diffusion Models \citep{ho2020denoising, song2020score}  has greatly advanced the field of motion generation guided by different prompts \cite{tevet2022human, zhang2024motiondiffuse}. \citet{chen2023executing} proposed a latent diffusion model for plausible human motion sequences. Physdiff \cite{yuan2023physdiff} introduce physical information into the human motion diffusion model (MDM) \cite{tevet2022human}. Diffusion models have also been found useful in a variety of other human motion generation tasks such as human-objects interaction \cite{xu2023interdiff}, objects to motion \cite{li2023object}, motion in-between \cite{cohan2024flexible}, crowd simulation \cite{ji2024text}, and two-person 
interactions \cite{tanaka2023role, liang2024intergen, wang2023intercontrol, ruiz2024in2in}.

ComMDM \cite{shafir2023human} adapts MDM \cite{tevet2022human} to perform two-person motion generation. Their results are restricted to the original single human motion distribution, which cannot produce diverse human interactions. To alleviate the shortage of data, InterGen \cite{liang2024intergen} captures a large two-person interaction dataset. They introduce two parallel transformer-based denoisers that explicitly share weights to generate two people simultaneously. However, their architecture limits their ability to expand to interactions involving multiple people. \citet{shan2024towards} propose a new multi-person dataset derived semi-automatically from images and videos using mesh reconstruction techniques. Their model suffers from low-quality estimated motion, and the motion length is sacrificed to increase the number of subjects in the generation. FreeMotion \cite{fan2024freemotion} presents a self-attention architecture to learn interaction information. Their framework generates the motion of each person in the interaction sequentially. Each newly generated motion is then concatenated as a condition for generating the motion of the next person, until all motions have been generated. 
While such a structure allows the model to generate multi-person interactions, 
they do not explore the complex interaction relationships in multi-person motions since every previous motion is concatenated in the same way. As a result, FreeMotion only generates simple and repetitive interactions, with everyone performing the same actions. 

By leveraging existing two-person models, our Graph-driven Interaction Sampling enables users to precisely control the interaction relationships between multiple characters, generating more diverse and plausible multi-person interaction motions \textit{without any retraining}. Additionally, the proposed optimization process further reduces the occurrence of intersections and other artifacts which further improves the interaction quality not only in multi-character, but also in two-character scenarios.

%% file: sec/3_Preliminaries.tex
\section{Preliminaries}
\label{sec:prelim}

\subsection{Diffusion models}
\label{sec:prelims-diffusion}

Denoising diffusion models \cite{ho2020denoising, sohl2015deep} define both a forward diffusion process and a reverse diffusion process. The forward diffusion process incrementally adds noise to real data $x_0$ over $T$ steps and is defined as
\begin{equation}
\centering
q\left ( x_t|x_0 \right )= \mathcal{N}\!\left ( x_t;\, \sqrt{\bar{\alpha}_t}x_0 ,\, \left ( 1- \bar{\alpha}_t\right )I\right)
\end{equation}
where 
$\bar{\alpha}_t= \prod_{i=1}^{t}\alpha_t$ for some constants $\alpha_t$ defining the noise schedule, $I$ is the identity matrix, and $t \in [0, T]$ is the timestep.

The reverse diffusion process aims to produce the desired data distribution from random noise $x_T \sim N\left (0,I  \right )$. This is achieved through an iterative denoising process facilitated by a neural network $m_\theta$ with parameters $\theta$.
More precisely, at timestep $t$, given conditioning information (e.g.~a text prompt) $c$, the reverse diffusion process can be represented as
\begin{equation}
p_\theta \left ( x_{t-1}|x_t,c \right )= \mathcal{N}\!\left ( x_{t-1} ;\, \mu _\theta \left ( x_t,t,c \right ) ,\, \left ( 1- \alpha_t\right )I\right)
\end{equation}
where $\mu _\theta \left ( x_t,t,c \right )$ is the estimated posterior mean. Following MDM \citep{tevet2022human}, we predict the original motion $x_0 = m_\theta(x_t,t,c)$, so $\mu _\theta $ is given by
\begin{equation}
\mu _\theta \left ( x_t,t,c \right ) = \frac{\sqrt{\bar{\alpha}_{t-1}}\beta _t}{1-\bar{\alpha}_{t}} x_0(x_t,t,c;\theta) + \frac{\sqrt{\alpha}_t (1-\bar{\alpha}_{t-1})}{1-\bar{\alpha}_{t}} x_t
\end{equation}
%
where $\beta_t = 1- \alpha_t$. 

\subsection{Two-person interaction diffusion model}
\label{sec:prelims-tpm}

We denote a single human motion as $x$, and a two-person interaction as a pair $\{x^i, x^j\}$, where $x^n = \big(x^n_{(k)}\big)_{k=1}^{L}$ and $L$ is the number of frames of $x^n$.
Our framework leverages existing models that can denoise one person's motion conditioned on another; specifically we can use two kinds of motion diffusion model.
First, we use two-person interaction diffusion models (e.g. InterGen~\cite{liang2024intergen})---at each diffusion timestep $t$,  these calculate $\{x^i_{t-1}, x^j_{t-1}\}$ given $\{x^i_{t}, x^j_{t}\}$ and text conditions $c$. Thus, when denoising ${x^i_t}$, we provide ${x^j_t}$ as a condition, and vice versa. As a result, two-person interaction generation can be regarded as \textit{a special case of conditional single-person motion generation}, where we condition on a still-noisy motion.
Second, we use reaction generation (e.g. ReMoS~\cite{ghosh2025remos})---these can be interpreted similarly, now with clean motion ${x^j_0}$ of one person serving as the condition when denoising ${x^i_t}$.

%% file: sec/4_method.tex
\section{Method}
\label{sec:method}

We present \textbf{Graph-driven Interaction Sampling}, a method for generating realistic multi-person interactions as illustrated in Fig.~\ref{fig:graph}. Our key idea is to recast the multi-person motion generation task as simultaneously generating multiple single-person motions, with each person's motion conditioned on another. This enables using existing two-person motion diffusion models, since two-person motion generation may be viewed as conditional single-person generation where each serves as a condition for the other. To realize this idea, our framework consists of two parts: a graph-based multi-person sampling scheme where a Pairwise Interaction Graph defines the decomposition of the multi-person motion, and a sampling optimization process that ensures coherent interactions between all characters.

Our graph-driven interaction sampling scheme reinterprets the conditional distributions learnt by the existing two-person models as factors in a pairwise energy-based model defined over a \textit{Pairwise Interaction Graph} (Sec.~\ref{sec:method-ig}). This graph mediates interactions between the motions of different persons, ensuring a coherent multi-person motion is sampled despite the two-person motion diffusion model only generating two persons each time. 

To ensure the multi-person interactions are globally coherent (e.g.~not exhibiting intersections), we propose interaction sampling optimization (Sec.~\ref{sec:method-ino}), which uses two new losses -- \textit{Proxemics loss} 
and \textit{Gauss Linking Integral (GLI) loss}. The Proxemics loss aims to maintain a distance between two people who are 
not interacting with each other in the Pairwise Interaction Graph. To further preserve the consistency of spatial relationships over time and reduce interpenetration of body parts, we introduce the GLI loss to penalize significant changes in spatial relationships in consecutive frames. 

During sampling, the user first defines a Pairwise Interaction Graph specifying pairwise relationships among $n$ persons. The scheme then starts $n$ denoising processes and generates motions conditioned on the pairwise relationship and interaction text. The interaction sampling optimization is applied at each denoising step.

\subsection{Graph-based Multi-person Sampling}
\label{sec:method-ig}

To enable sampling motion involving three or more people without training a new model, we introduce a novel graph-driven interaction sampling scheme which can be used with pre-trained two-person interaction synthesis models such as InterGen~\cite{liang2024intergen} and in2IN~\cite{ruiz2024in2in}.

\begin{figure} [htb]
  \centering
  \begin{subfigure}{0.45\linewidth}
    \includegraphics[width=\linewidth]{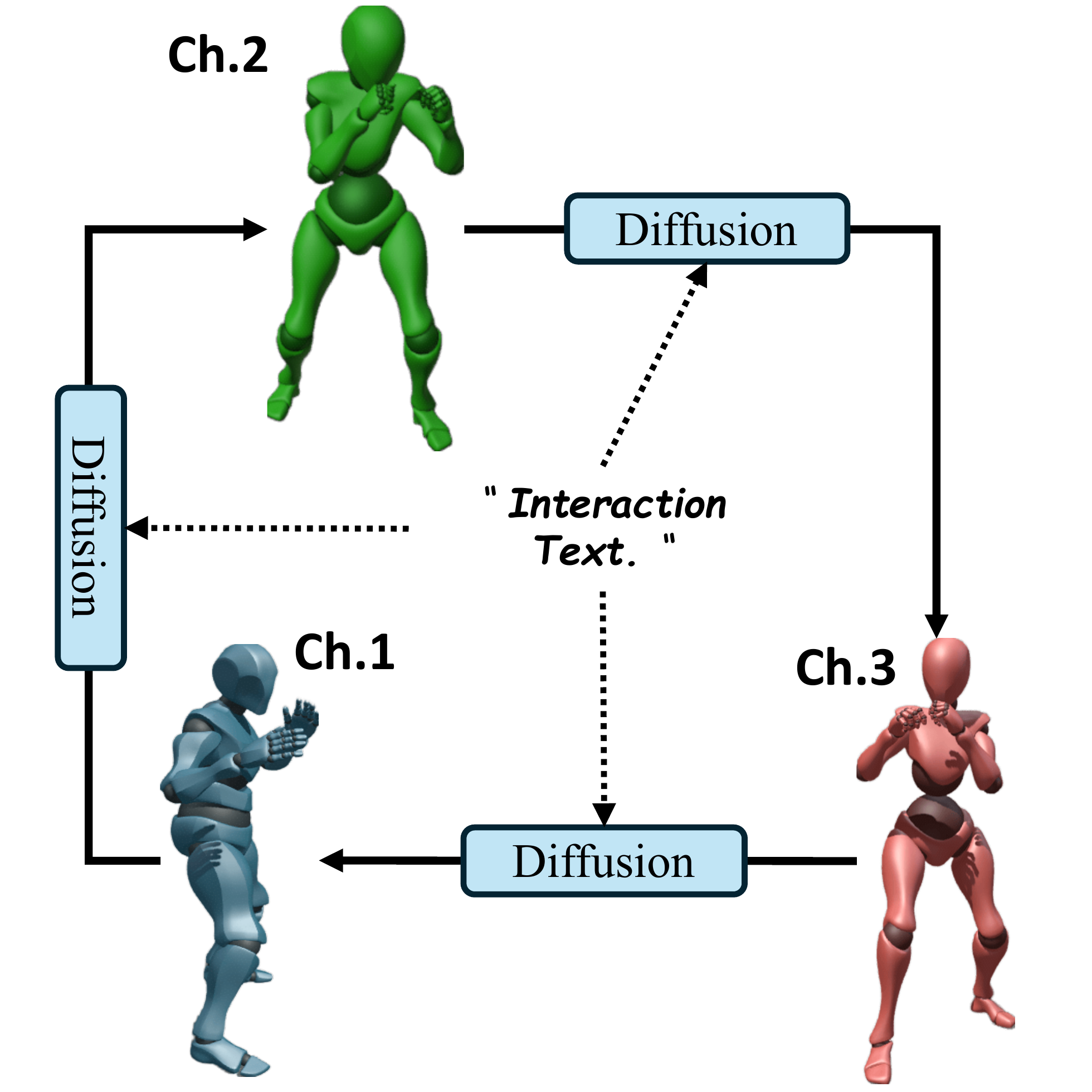}
    \caption{cyclic}
    \label{fig:graph_1}
  \end{subfigure} 
  \begin{subfigure}{0.45\linewidth}
    \includegraphics[width=\linewidth]{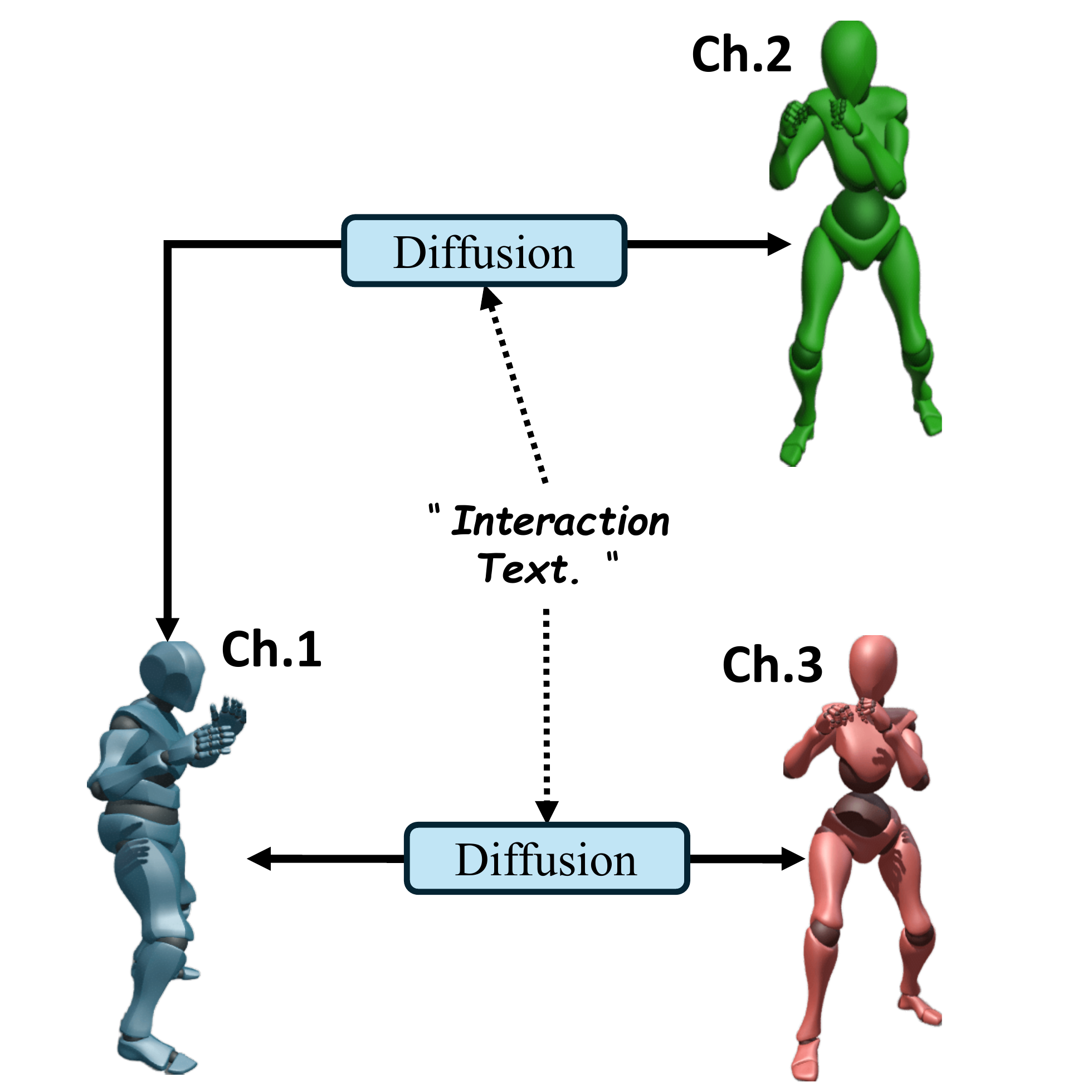}
    \caption{``2 versus 1''}
    \label{fig:graph_2}
  \end{subfigure} 
  \caption{Examples of interaction graph for decomposing a multi-person interaction into coupled pairwise interactions.}
  \label{fig:graph}
\end{figure}

We use a \textit{Pairwise Interaction Graph} to decompose the multi-person interaction into coupled pairwise interactions. Examples in Fig.~\ref{fig:graph_1} and Fig.~\ref{fig:graph_2} illustrate how three-person interactions can be decomposed according to different graph structures; each person's motion is conditioned on the motion of the one that an arrow points from and the interaction text. 
In Fig.~\ref{fig:graph_1}, the three individuals are cyclically conditioned on each other and sampled in parallel. 
Fig.~\ref{fig:graph_2} illustrates that two individuals attack one another simultaneously. 
Thus, our framework enables superior precision in controlling the generated interactions. Users are free to specify a Pairwise Interaction Graph that captures the most important dependencies for their specific task prior to sampling. 

More formally, the Pairwise Interaction Graph is a directed graph where each vertex corresponds to the motion of one person, and directed edges indicate conditional dependencies between motions. 
Each conditional dependency reflects how one person is affected by one other, consistent with how current two-person diffusion models are trained (Sec.~\ref{sec:prelim}). 
We can interpret the conditional probabilities $p(x_j | x_i)$ learnt by the existing two-person diffusion model as factors in a factor-graph model, whose structure is an undirected version of the Pairwise Interaction Graph. This model defines a joint probability distribution over the multi-person motion. Specifically, the probability density for the motion $(x^1, x^2,\ldots, x^n)$ of $n$ persons interacting is defined as:
\begin{equation}
p(x^1, x^2,\ldots, x^n) = \frac{1}{Z} \prod_{(i,j) \in \mathrm{edges}} \phi_{i\rightarrow j}(x_i, x_j)
\label{eq:multi-pdf}
\end{equation}
where $Z$ is a normalizing constant and the factor between persons $i$ and $j$ is set to $\phi_{i\rightarrow j}(x_i, x_j) = p(x_j \,|\, x_i, c)$, i.e.~the conditional probability of the $j$\textsuperscript{th} person's motion given the $i$\textsuperscript{th} and the text condition $c$. 

Intuitively, the joint distribution (Eq.~\ref{eq:multi-pdf}) assigns high probability density to interacting motions where each character's motion is itself plausible, and also coherent with that of the characters on which it is conditioned.
Provided the graph is connected, the set of pairwise dependencies encoded by the factors induces correlations between the motions of all persons, yielding coherent multi-person interactions. Unfortunately, the distribution (Eq.~\ref{eq:multi-pdf}) is intractable to sample directly for general Pairwise Interaction Graphs.
We therefore propose a straightforward but effective approximate sampling process, which we show yields high-quality and coherent multi-person interactions.
Specifically, we generate the $n$ motion sequences $(x^1, x^2,\ldots, x^n)$ by simultaneously running $n$ denoising processes, each conditioned on the motion of the person indicated by the graph, and with other conditions like text prompts. As mentioned in Sec.~\ref{sec:prelims-tpm}, at each timestep $t$, we do one denoising step to $t-1$ for each motion conditioned on ${x^n_t}$ of others at time $t$, then update them all, and the process is repeated in the next timestep.

For example, using InterGen and a fixed graph, we start from $n$ pure noise samples $(x^1_T, x^2_T,\ldots, x^n_T)$. During each diffusion step, an independent denoising calculation is performed for each conditioned pair (i.e.~edge of the graph). More specifically, given predicted clean motion ${\hat{x}^{(j\rightarrow i)}_0}$ and ${\hat{x}^{(k\rightarrow i)}_0}$ for person $i$ conditioned respectively on persons $j$ and $k$, we calculate the final $\hat{x}^i_0$ for the timestep by averaging them. Then, we take a denoising step towards this as usual.

\subsubsection{Time-varying Pairwise Interaction Graphs.}

\begin{figure} [htb]
  \centering
  \begin{subfigure}{0.45\linewidth}
    \includegraphics[width=\linewidth,trim={0 0 0 3cm},clip]{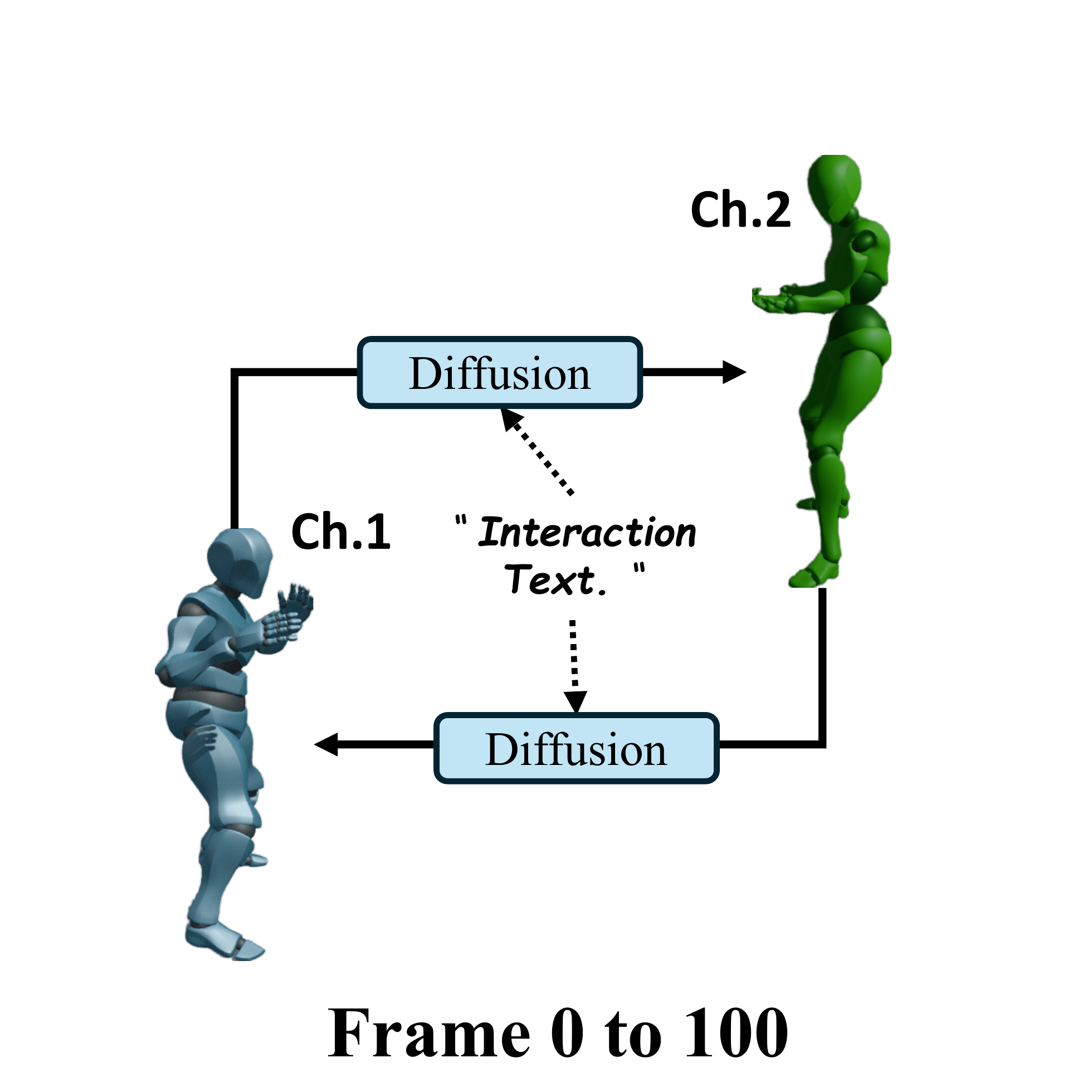}
    \caption{Ch.1 interacts with Ch.2 in first 100 frames}
    \label{fig:graph_31}
  \end{subfigure} 
  \hfill     
  \begin{subfigure}{0.45\linewidth}
    \includegraphics[width=\linewidth,trim={0 0 0 3cm},clip]{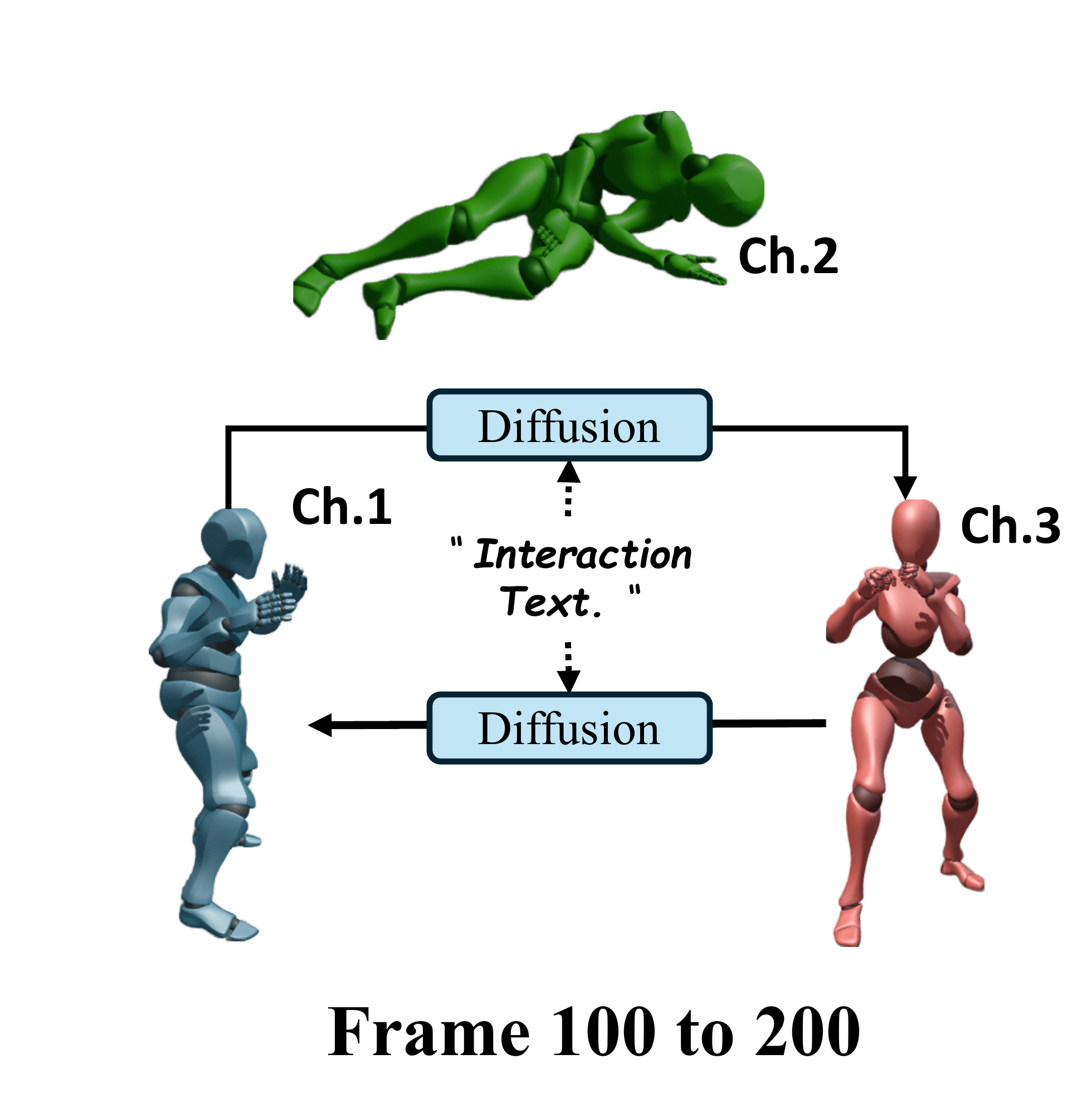}
    \caption{Ch.1 then interacts with Ch.3 in last 100 frames}
    \label{fig:graph_32}
  \end{subfigure}

  \caption{An example of time-varying interaction graph with character 1 is conditioned on both characters 2 and 3.}
  \label{fig:time-varying}
\end{figure}

In complex interactions, timing is crucial in defining relationships between characters. 
In daily life, we often interact with specific individuals or smaller sub-groups separately, rather than engaging with an entire group all at once. To model such scenarios, we introduce \textit{time-varying} Pairwise Interaction Graphs.
For example, the blue character attacks the green character in the first half of the interaction (Fig.~\ref{fig:time-varying}(a)), then gets attacked by the red character in the second half (Fig.~\ref{fig:time-varying}(b)). The time-varying Pairwise Interaction Graph therefore initially has edges between the green and blue characters mediating their interaction, before changing to edges between the red and blue. 
We now write the factors as $\phi_{i(f)\rightarrow j(f)}(x_{i(f)}, x_{j(f)}) = p(x_{j(f)} \,|\, x_{i(f)}, c)$, where $f$ specifies a subsequence of frames. 

To produce temporally coherent motion for each character, during each denoising step, we consider all factors conditioning that character, finding clean $\hat{x}^{(i(f)\rightarrow j(f))}_0$ for each. For frames where several factors overlap temporally, we average the clean predictions due to each relevant factor to give the final $\hat{x}^i_0$, similar to the process for multiple conditioning characters described above.

\subsection{Interaction Sampling Optimization}
\label{sec:method-ino}
When synthesizing close interactions between characters, artifacts such as interpenetrations of body parts can be seen in the interactions generated using existing two-person interaction synthesis models (see Sec.~\ref{sec:2ppl_exp}), not to mention when multiple characters (see Sec.~\ref{sec:mppl_exp}) are engaged in close interactions. Here, we propose Interaction Sampling Optimization to maintain the quality of interactions while reducing interpenetrations in close interaction. In the optimization step, two new loss terms are designed to alleviate the aforementioned problems. 

\subsubsection{Optimization Losses}
\label{sec:method-ino-loss}

\paragraph{Proxemics Loss} Although graph-driven interaction sampling generates different patterns of multi-person interactions, some heavy interpenetration cannot be avoided when the two colliding characters are unconnected (i.e. non-interacting) or distant in the Pairwise Interaction Graph. 
Two characters that are not neighbors in the graph will lack correlations, as our sampling process fails to induce any between them. 
Therefore, during the optimization process, we introduce a new Proxemics Loss to ensure that unconnected characters do not exhibit significant overlapping. For example, as shown in Fig.~\ref{fig:graph_2}, Ch.2 and Ch.3 are unconnected 
in the graph. Specifically, the bounding box of each character is calculated and we take the overlap volume of pairs of bounding boxes as the loss, punishing the overlaps between unconnected characters. The root distance can be added as a penalty as well when we need to push characters away from each other. We traverse the entire Pairwise Interaction Graph and sum over the proxemics loss of all unconnected pairs.

\begin{figure}[htb]
\centering
     \begin{subfigure}[b]{0.20\linewidth}
         \centering
         \includegraphics[width=\textwidth]{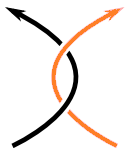}
         \caption{\footnotesize 0.5<GLI<1}
     \end{subfigure}    \hspace{2mm} 
     \begin{subfigure}[b]{0.20\linewidth}
         \centering
         \includegraphics[width=\textwidth]{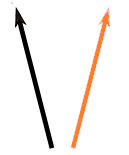}
         \caption{\footnotesize 0<|GLI|<0.5}
     \end{subfigure}      \hspace{2mm}
     \begin{subfigure}[b]{0.20\linewidth}
         \centering
         \includegraphics[width=\textwidth]{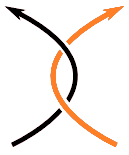}
         \caption{\footnotesize -1<GLI<-0.5}
     \end{subfigure}          \hspace{2mm} 
     \begin{subfigure}[b]{0.20\linewidth}
         \centering
         \includegraphics[width=\textwidth]{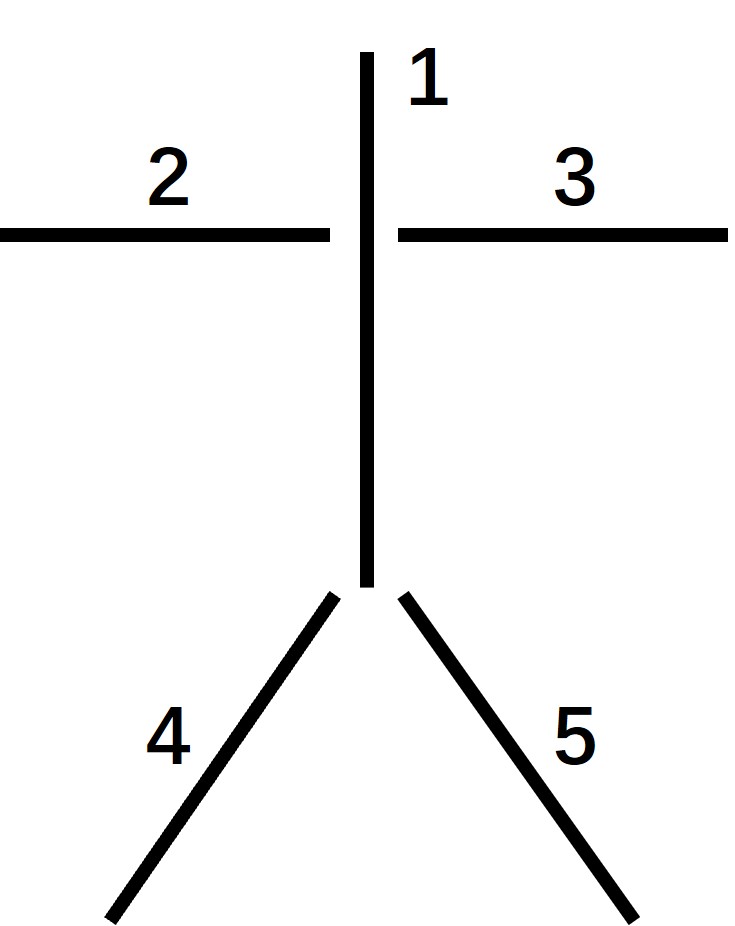}
         \caption{}
     \end{subfigure}
\caption{(a)-(c) Examples of GLI values computed from different configurations of a pair of serial chains representing the body parts. (d) Representing a skeletal pose by 5 serial chains.}
\label{fig:SerialChain}
\end{figure}

\paragraph{Gauss Linking Integral (GLI) loss} Inspired by the success of using GLI as soft constraints in an optimization-based approach \cite{Ho:TopoCoor} for generating collision-free close interactions, we propose monitoring the pairwise GLIs of closely interacting body parts. Given two directed curves $\gamma_1$ and $\gamma_2$, the GLI~\citep{Pohl:GLI} can be calculated by Eq. \ref{eq:gli} by integrating along the two curves:
\begin{equation}
\label{eq:gli}
G(\gamma_1, \gamma_2) = \frac{1}{4\pi} \int_{\gamma_1} \int_{\gamma_2} \frac{{dr_1} \times {dr_2} \cdot (r_1 - r_2)}{{\|r_1-r_2 \|}^3}
\end{equation}
where $r1$ and $r2$ are the parameterization of $\gamma_1$ and $\gamma_2$ (i.e. 3D coordinates along the curves), $\times$ and $\cdot$ are cross-product and dot-product operators, respectively.  GLI is a signed scalar value which computes the average number of crossings from all viewing directions of the two curves, such as Figure~\ref{fig:SerialChain}~(a-c). Please refer to the Supplementary Materials for the discretization of Eq.~\ref{eq:gli} and the analytical solution for GLI calculation. Here, we represent each character as 5 serial chains based on the 3D joint positions as shown in Fig.~\ref{fig:SerialChain}~(d) and calculate the GLI from each pair of serial chains extracted from two characters.

By ensuring the GLIs change smoothly over time, implausible motions such as interpenetrations can be avoided since the GLI will change significantly (e.g. changing the sign of the scalar value) when interpenetration happens.

Specifically, during the optimisation, we compute the difference between the GLIs computed on two consecutive frames from each pair of chains. If the difference is above an empirically determined threshold of $0.4$, it will be added to the loss to discourage the optimisation from picking/sampling such an interaction. We calculate the summed GLI loss for every directly interacting character.

{\begin{table*}[!htb]
  \caption{Quantitative comparisons on the InterHuman test set. ± indicates the 95\% confidence interval. Bold face indicates the best result}
  \label{tab:two-person}
  \centering
  \begin{tabular}{lccccccc}
    \toprule
    & R Precision (Top3)$\uparrow$ & FID$\downarrow$ & MM Dist $\rightarrow$ & Diversity $\uparrow$ & PeneBone (m) $\downarrow$ & Skating ratio $\downarrow$ & Jitter $\downarrow$\\
    \midrule
    Real & $0.705^{\pm{0.005}}$ & $0.291^{\pm{0.006}}$ & $3.782^{\pm{0.002}}$ & $7.782^{\pm{0.032}}$ & $0.256^{\pm{0.006}}$ & $0.069^{\pm{0.001}}$ & $7.083^{\pm{0.008}}$\\
    \midrule
    FreeMotion & $0.544^{\pm{0.006}}$ & {\boldmath$6.740^{\pm{0.130}}$} & {$3.848^{\pm{0.002}}$} & {$7.828^{\pm{0.130}}$} & {$1.443^{\pm{0.083}}$} & $0.074^{\pm{0.001}}$ & $4.506^{\pm{0.128}}$\\
    InterGen & {\boldmath$0.663^{\pm{0.005}}$} & {$6.883^{\pm{0.124}}$} & {\boldmath$3.797^{\pm{0.002}}$} & {$7.807^{\pm{0.026}}$} & $0.899^{\pm{0.046}}$ & {\boldmath$0.071^{\pm{0.001}}$} & $4.292^{\pm{0.028}}$\\
    Ours & $0.638^{\pm{0.013}}$ & $6.890^{\pm{0.146}}$ & {$3.819^{\pm{0.002}}$} & {\boldmath$7.853^{\pm{0.026}}$} & {\boldmath$0.211^{\pm{0.010}}$} & $0.072^{\pm{0.001}}$ & {\boldmath$4.282^{\pm{0.115}}$}\\
    \bottomrule
  \end{tabular}
\end{table*}}

\subsubsection{Diffusion Posterior Sampling Optimization}

Without adjusting the model, the posterior sampling optimization algorithm enables control over the generated results to satisfy constraint requirements. In our task, we follow a setup similar to Diffusion Posterior Sampling (DPS)~\cite{chung2023diffusion}.
DPS calculates reconstruction errors between predictions and observations from the posterior samples obtained through prediction and uses these scores to adjust the sampling direction at each step.
Instead of a reconstruction error as in DPS, we use the GLI and proxemics losses defined above. 
During each diffusion sampling step, we calculate these losses based on the current sample $\hat{x}_{t-1}$
and iteratively update it by gradient descent. 
In a scenario where two people $i$ and $j$ are interacting, 
this leads to an update for the $i$\textsuperscript{th} person in the form
\begin{equation}
\label{eq:dps}
x_{t-1}^{i} \leftarrow x_{t-1}^{i} - \lambda_{t}\nabla_{x_{t}^{i}}(L_{G}(\hat{x}_{0}^{i}, \hat{x}_{0}^{j}))
\end{equation}
%
where $L_{G}$ is the GLI loss.
After adding a third person $k$ who is not adjacent to $i$ this becomes:
\begin{equation}
\label{eq:dps2}
x_{t-1}^{i} \leftarrow x_{t-1}^{i} - \lambda_{t}\nabla_{x_{t}^{i}}(L_{G}(\hat{x}_{0}^{i}, \hat{x}_{0}^{j}) + L_{P}(\hat{x}_{0}^{i}, \hat{x}_{0}^{k}))
\end{equation}
where $L_{P}$ is the Proxemics Loss. 


%% file: sec/5_experiments.tex
\section{Experiments}
\label{sec:experiments}
As we leverage existing two-person models, we first introduce the two-person dataset and baseline we use for multi-character interactions. Then, we evaluate the effectiveness of our optimization method in reducing the penetration in two-person interactions. Lastly, we demonstrate our multi-character results and ablation studies. 

\paragraph{Datasets.}
The InterHuman \cite{liang2024intergen} dataset consists of 6,022 recordings of two-person interactions with corresponding text annotations, at a framerate of 30~Hz, is used in the evaluation. SMPL-H~\cite{romero2017embodied} is used to represent the human shape. We selected a small close interaction subset from the InterHuman test set with around 20,000 frames to evaluate our method.

\paragraph{Metrics.}
We calculate R-Precision, FID, Diversity, multi-modal distance (MM Dist) to measure text-to-motion performance following \cite{liang2024intergen}, and PeneBone (Penetration Bone Level), Skating ratio, and Jitter for artifacts.
\textit{R-Precision} assesses how closely generated motions align with their conditioning text, considering the prompt as successful if it ranks within the top-$N$ closest matches out of 95 sampled negatives.
\textit{FID (Fr\'{e}chet Inception Distance)} measures the distributional distance between generated and real motions, \textit{Diversity} captures motion variance, and \textit{MM Dist} calculates the mean L2 distance between paired text and motion features, 
\textit{PeneBone} measures the collision depth between characters' skeletons by modeling each `bone' using a box with a radius of 2cm. %
\textit{Skating ratio} \cite{guo2022generating} measures the percentage of frames where either foot exhibits noticeable sliding (2.5 cm) while remaining in contact with the ground (foot height < 5 cm). 
\textit{Jitter} \cite{karunratanakul2023guided} reflects motion smoothness, computed as the mean acceleration change across all joints. For close and multi-person interactions, we introduce two new metrics \textit{Contact} (Contact points per frame) and \textit{CFrame} (Contact frame rate) to represent the collision magnitude between characters when the generated motions are fitted to the SMPL model. Lastly, \textit{Pair-FID} is the FID of all interaction pairs in the multi-person interactions.

\paragraph{Baselines.}
For two-person interaction generation, we use InterGen \cite{liang2024intergen} as the baseline. As our graph-driven interaction sampling is plug-and-play and model-agnostic, we use InterGen as the backbone of our method for multi-person as well. 
For multi-person generation, we compare with FreeMotion \cite{fan2024freemotion}, which is the only other method supporting generating a variable number of multi-person interactions. FreeMotion generates multi-person interaction sequentially, with all generated motions concatenated to an additional self-attention block. This self-attention block will finally be the condition of the next single-person motion generation. Finally, to demonstrate the versatility of our method, we also used in2IN \cite{ruiz2024in2in} as a backbone to generate multi-person interactions, showing that our approach can be seamlessly plugged into various other models. The results are shown in Table~\ref{tab:ablation}. 

{\begin{table}[!htb]
  \caption{Quantitative results on a subset with mostly close interactions}
  \label{tab:closeinter}
  \centering
  \begin{tabular}{lcccc}
    \toprule
    & {FID} & PeneBone (m) & Contact & CFrame(\%)\\
    \midrule
    Real & $10.350$ & $0.583$ & - & -\\
    \midrule
    InterGen & {\boldmath$30.234$} & $2.335$ & $136.657$ & $30.646 $\\
    Ours & {$31.140$} & {\boldmath$0.525$} & {\boldmath$34.693$} & {\boldmath$10.788$}\\
    \bottomrule
  \end{tabular}
\end{table}}

\subsection{Two-person interactions} \label{sec:2ppl_exp}
Table~\ref{tab:two-person} shows that the quality of the interactions generated by our method is comparable to the baselines, while we significantly reduced 76.53\% and 85.38\% of the penetration depth over InterGen~\cite{liang2024intergen} and FreeMotion~\cite{fan2024freemotion}, respectively. This highlights our method can effectively remove artifacts in close interactions which is comparable to the real motions in the dataset. Our method is more effective in handling closer interactions as demonstrated in Table~\ref{tab:closeinter}, with a significant reduction in penetration depth and number of contacts by 77.52\% and 74.61\% over InterGen. Qualitative results can be found in Fig.~\ref{fig:togli} and the video demo.

\subsection{Multi-person interactions} \label{sec:mppl_exp}
For multi-person interaction generation, we focus on qualitative evaluation due to the lack of ground truth data for numerical analysis. Nevertheless, some quantitative evaluations are presented in Table~\ref{tab:multi}. It can be seen that our method outperformed FreeMotion in all metrics over all settings. Cyclic graph is used for fair comparison. 

For Pair-FID, our method is robust to different numbers of characters while the quality of the motion generated from FreeMotion degrades when there are more characters in the scene. The Pair-FID obtained by FreeMotion is 1.70 and 1.60 times larger than ours with 4 and 5 characters in the scene, respectively. We believe this can potentially be caused by the repetitive motions of individual characters generated by FreeMotion which deviates from the natural interaction patterns in the ground truth data. 
Again, our method demonstrated superior performance in reducing interpenetration in the synthesized close interactions, with the results obtained using FreeMotion being 3.67 to 20.25 times higher in the PeneBone metric. 

For the qualitative analysis, 
Fig.~\ref{fig:demo} (a) and (b) show the screenshots of fighting interactions. The results demonstrate that our generated motion successfully avoided the severe interpenetration between the characters as shown in the motion generated using FreeMotion. 
The second example (Fig.~\ref{fig:demo} (d)) shows that our model allows more personal space around each character while dancing to avoid collisions. On the other hand, FreeMotion (Fig.~\ref{fig:demo} (c)) fails to deal with interactions involved with more characters. 
We further demonstrate the flexibility of our graph-driven interaction sampling approach by showing a wide range of interactions with 10-14 characters generated using different Pairwise Interaction Graphs (Fig.~\ref{fig:teaser} and \ref{fig:more_multi}).  Readers are encouraged to view the video demonstration in the supplementary materials to further evaluate the quality of the interactions generated using our proposed method.

{\begin{table}[!htb]
  \caption{Quantitative comparisons on multi-person interactions with 3 to 5 characters.}
  \label{tab:multi}
  \centering
  \resizebox{\linewidth}{!}{
  \begin{tabular}{@{}llcccc@{}}
    \toprule
    & & {Pair-FID} & PeneBone (m) & Contact & CFrame(\%)\\
    \midrule
    \parbox[t]{1mm}{\multirow{2}{*}{\rotatebox[origin=c]{90}{n=3}}} & FreeMotion & $81.621$ & $5.567$ & $271.518$ & $57.175$\\
    & Ours & {\boldmath$72.294$} & {\boldmath$1.515$} & {\boldmath$94.306$} & {\boldmath$23.809$}\\
        \midrule
    \parbox[t]{1mm}{\multirow{2}{*}{\rotatebox[origin=c]{90}{n=4}}} & FreeMotion & $113.907$ & $6.137$ & $332.103$ & $59.825$\\
    & Ours & {\boldmath$66.817$} & {\boldmath$0.303$} & {\boldmath$27.298$} & {\boldmath$13.397$} \\
            \midrule
    \parbox[t]{1mm}{\multirow{2}{*}{\rotatebox[origin=c]{90}{n=5}}} & FreeMotion & $106.742$ & $6.225$ & $303.967$ & $60.924$\\
    & Ours & {\boldmath$61.742$} & {\boldmath$0.456$} & {\boldmath$29.822$} & {\boldmath$10.027$}\\
    \bottomrule
  \end{tabular}%
}
\end{table}}

\subsection{Ablation}
\label{sec:experiments-ablation}
To understand the contribution of each component to the final performance, 
an ablation study is conducted and the results are presented in Table~\ref{tab:ablation}. Note that when $n=3$, Proxemics Loss will not be applied due to limited overlapping between characters. The results indicate having the loss term(s) (denoted by `Ours' with InterGen as backbone) can effectively reduce the penetration depths and number of contacts in the motion by 42.94\% and 29.94\% when $n=3$, 96.33\% and 91.60\% when $n=4$, and 87.80\% and 83.72\% when $n=5$, respectively. Note that our method introduces more foot skating; this may be caused by the normalisation of training data.

We further demonstrate the versatility of our model by using in2In~\cite{ruiz2024in2in} (denoted by `Ours (in2IN)' in Table~\ref{tab:ablation}) as the backbone. The penetration depths and number of contacts are comparable to those obtained using InterGen. 
We also conducted experiments on using MDM and FreeMotion as our backbone. Qualitative results can be found in Figure~\ref{fig:more_multi_2} and the demo video.

{\begin{table}[!htb]
  \caption{Quantitative results from the ablation study}  \label{tab:ablation}
  \centering
  \resizebox{\linewidth}{!}{
  \begin{tabular}{@{}ll@{}c@{\hskip 4pt}c@{\hskip 4pt}c@{\hskip 4pt}c@{\hskip 4pt}c@{}}
    \toprule
    & & PBone (m) & Contact & CFrame(\%) & Skating & Jitter\\
    \midrule
    \parbox[t]{1mm}{\multirow{3}{*}{\rotatebox[origin=c]{90}{n=3}}} & w/o GLI & $2.655$ & $134.614$ & $34.064$ & {\boldmath$0.074$} & $2.982$\\
    & Ours & {$1.515$} & {\boldmath$94.306$} & {$23.809$}  & $0.099$ & $2.930$\\
    \cline{2-7}
    & Ours (in2IN) & {\boldmath$1.004$} & {$98.084$} & {\boldmath$22.603$} & $0.127$ & $3.685$\\
        \midrule
    \parbox[t]{1mm}{\multirow{5}{*}{\rotatebox[origin=c]{90}{n=4}}} & w/o Proxemics & $4.919.$ & $192.381$ & $27.619$ & $0.079$ & $3.065$\\
    & w/o GLI & $0.753$ & $53.442$ & $15.778$ & $0.110$ & $2.913$\\
    & w/o both & $8.255$ & $285.836$& $38.428$ & {\boldmath$0.074$} & $3.027$\\
    & Ours & $0.303$ & $27.298$& $13.397$ & $0.087$ & $2.996$\\
    \cline{2-7}
    & Ours (in2IN) & {\boldmath$0.222$} & {\boldmath$24.002$}& {\boldmath$9.159$} & $0.133$ & $3.550$\\
            \midrule
    \parbox[t]{1mm}{\multirow{5}{*}{\rotatebox[origin=c]{90}{n=5}}} & w/o Proxemics & $4.518$ & $213.767$ & $35.762$ & $0.078$ & $3.146$\\
    & w/o GLI & {$0.708$} & $49.483$ & $11.873$ & $0.094$ & $2.805$\\
    & w/o both & $3.738$ & $183.152$ & $34.286$ & {\boldmath$0.073$} & $3.003$\\
    & Ours & {$0.456$} & {\boldmath$29.822$} & $10.027$ & $0.079$ & $2.848$\\
    \cline{2-7}
    & Ours (in2IN) & {\boldmath$0.415$} & {$34.886$}& {\boldmath$8.838$}  & $0.119$ & $3.454$\\
    \bottomrule
  \end{tabular}}
\end{table}}

In addition, we compared the performance of the GLI loss against a simple contact loss which measures the overlapping volume between the rectangular bounding boxes in Table~\ref{tab:GLILoss}. We find suitable loss weights such that the two losses result in a similar level of penetration depth on the whole test set (PeneBone-all) for fair comparison. While the simple contact loss yields a better R-precision value, the penetration depth (PeneBone-close) is worse than the GLI loss on the close interaction subset (used in Sec~\ref{sec:2ppl_exp}).

{\begin{table}[!htb]
  \caption{Ablation study of GLI loss}
  \label{tab:GLILoss}
  \centering
  \resizebox{\linewidth}{!}{
  \begin{tabular}{@{}cccc@{}}
    \toprule
     \(\lambda\)  GLI/simple & R-precision (Top3) & PeneBone-all & PeneBone-close\\
    \midrule
    $0.3/3$ & $0.656/${\boldmath$0.671$} & $0.453/0.401$ & {\boldmath$0.998$}$/1.278$ \\
    $0.5/5$ & $0.653/${\boldmath$0.667$} & $0.313/0.323$ & {\boldmath$0.875$}$/0.916$\\
    $0.7/7$ & $0.641/${\boldmath$0.654$} & $0.274/0.264$ & {\boldmath$0.701$}$/0.845$\\
    $1.0/10$ & $0.638/${\boldmath$0.655$} & $0.211/0.231$ & {\boldmath$0.525$}$/0.679$ \\
    \bottomrule
  \end{tabular}%
}
\end{table}}

%% file: sec/6_discussion.tex
\section{Conclusion}
\label{sec:discussion}

In summary, we introduce Graph-Driven Interaction Sampling, a novel framework for generating multi-person interactions. By leveraging prior knowledge from the existing two-person motion diffusion model, as well as decomposing complex multi-person interactions into paired two-person interactions, our method can generate various multi-person interactions without using any multi-person interaction dataset or retraining a new model. We further introduced the new Proxemics and GLI losses to reduce intersections between characters in the scene. The effectiveness of our method is demonstrated in the experimental results.

\paragraph{Limitations and future work}
Currently, the interaction graph is given by the user, which maximizes controllability. In the future, we are interested in providing suggestions to the user on the trajectories of each person and the pairwise interaction graphs. This can potentially be achieved by an LLM with properly designed prompts. In addition, as our current method cannot achieve totally penetration-free, we plan to utilize physics simulation to further reduce artifacts such as intersections in complex interactions with a lot of characters.

Furthermore, our method is totally based on existing two-person models, and cannot capture relationship between nodes that are not directly connected. We can improve it by incorporating ‘auxiliary conditioning’ on other characters (e.g. quantize the surroundings and encode the occupancy to inform the proxemic loss calculation).

%% file: sec/figure-only.tex
\begin{figure*}
  \setlength{\belowcaptionskip}{7pt}
  \setlength{\abovecaptionskip}{0pt}
  \centering
    \begin{subfigure}{0.45\linewidth}
    \includegraphics[width=\linewidth,trim={0 1.0cm 0 1.5cm},clip]{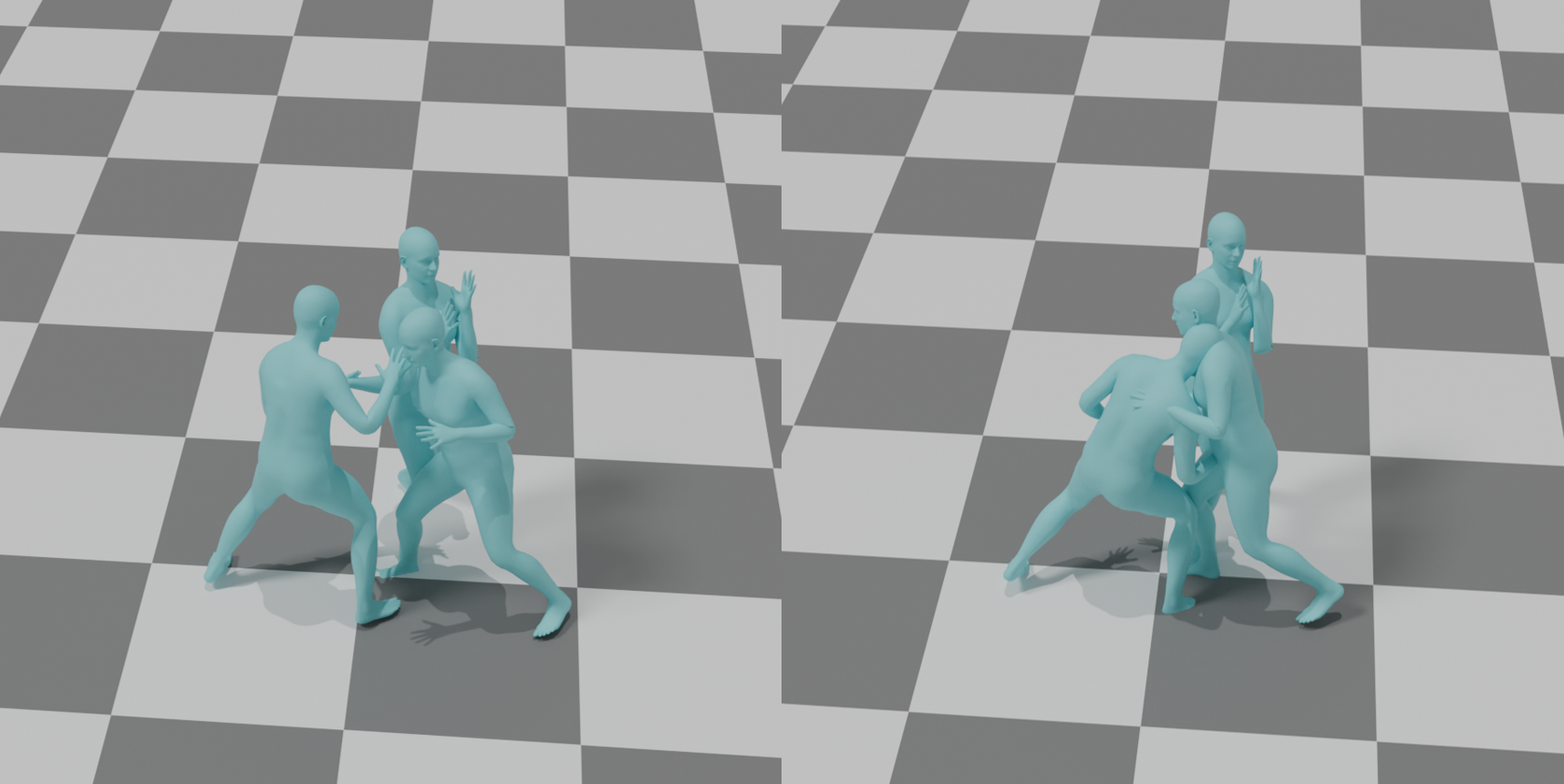}
    \caption{3-person fighting generated by FreeMotion}
    \label{fig:demo_afree}
  \end{subfigure}
  \begin{subfigure}{0.45\linewidth}
    \includegraphics[width=\linewidth,trim={0 1.0cm 0 1.5cm},clip]{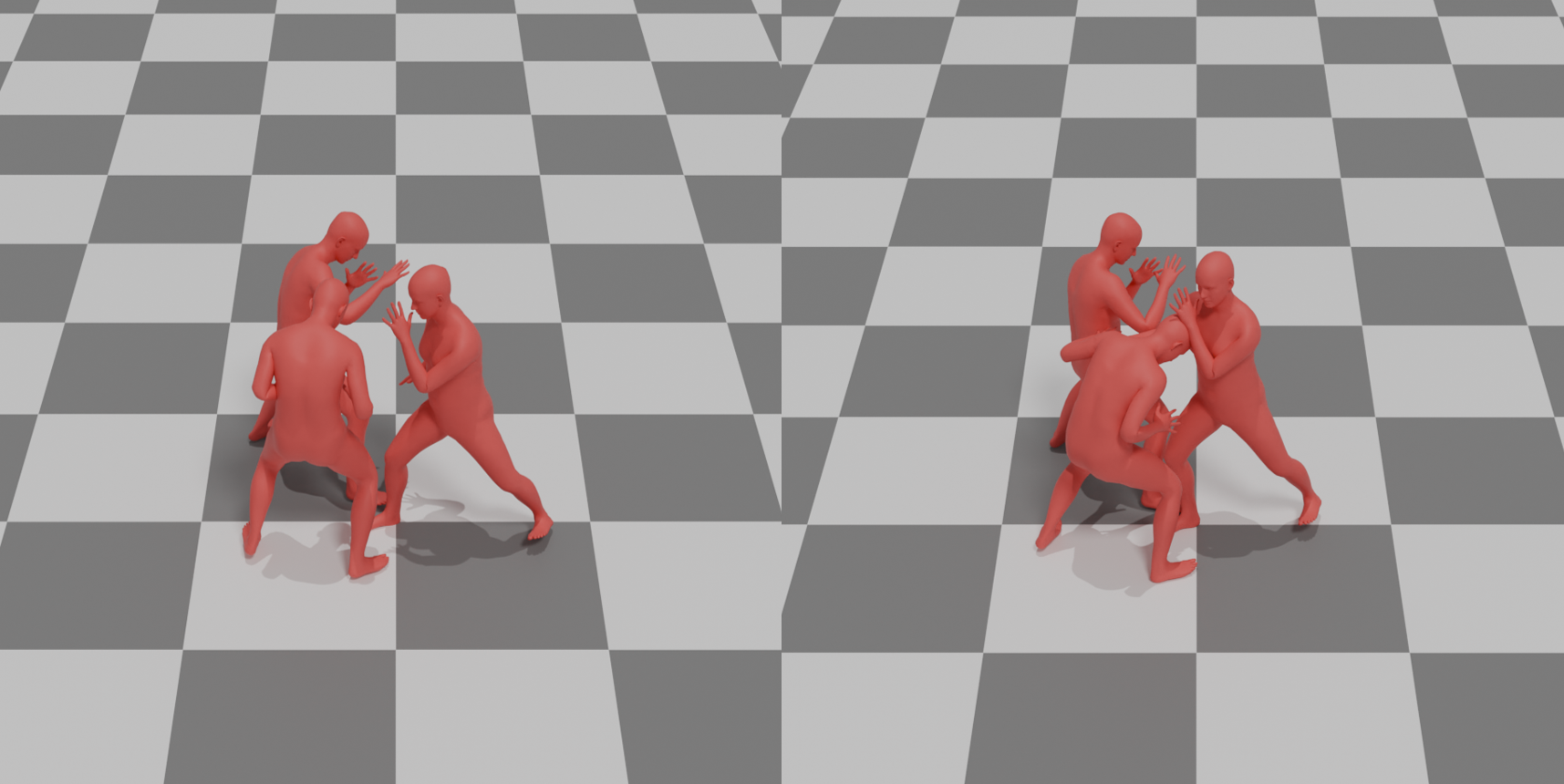}
    \caption{3-person fighting generated by ours}
    \label{fig:demo_a}
  \end{subfigure}
  \begin{subfigure}{0.45\linewidth}
    \includegraphics[width=\linewidth,trim={0 1.0cm 0 1.25cm},clip]{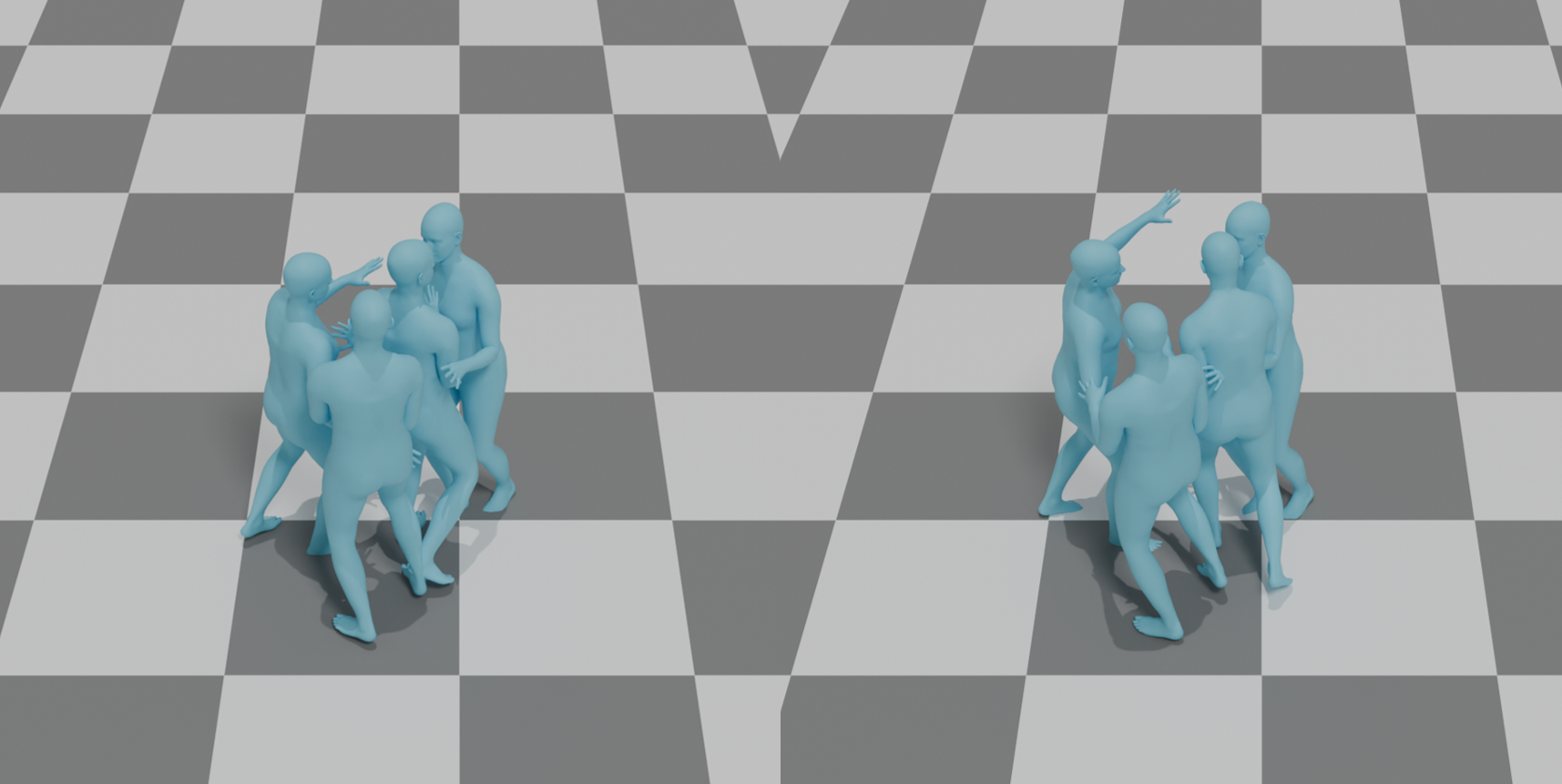}
    \caption{4-person dancing generated by FreeMotion}
    \label{fig:demo_bfree}
  \end{subfigure}
    \begin{subfigure}{0.45\linewidth}
    \includegraphics[width=\linewidth,trim={0 1.0cm 0 1.25cm},clip]{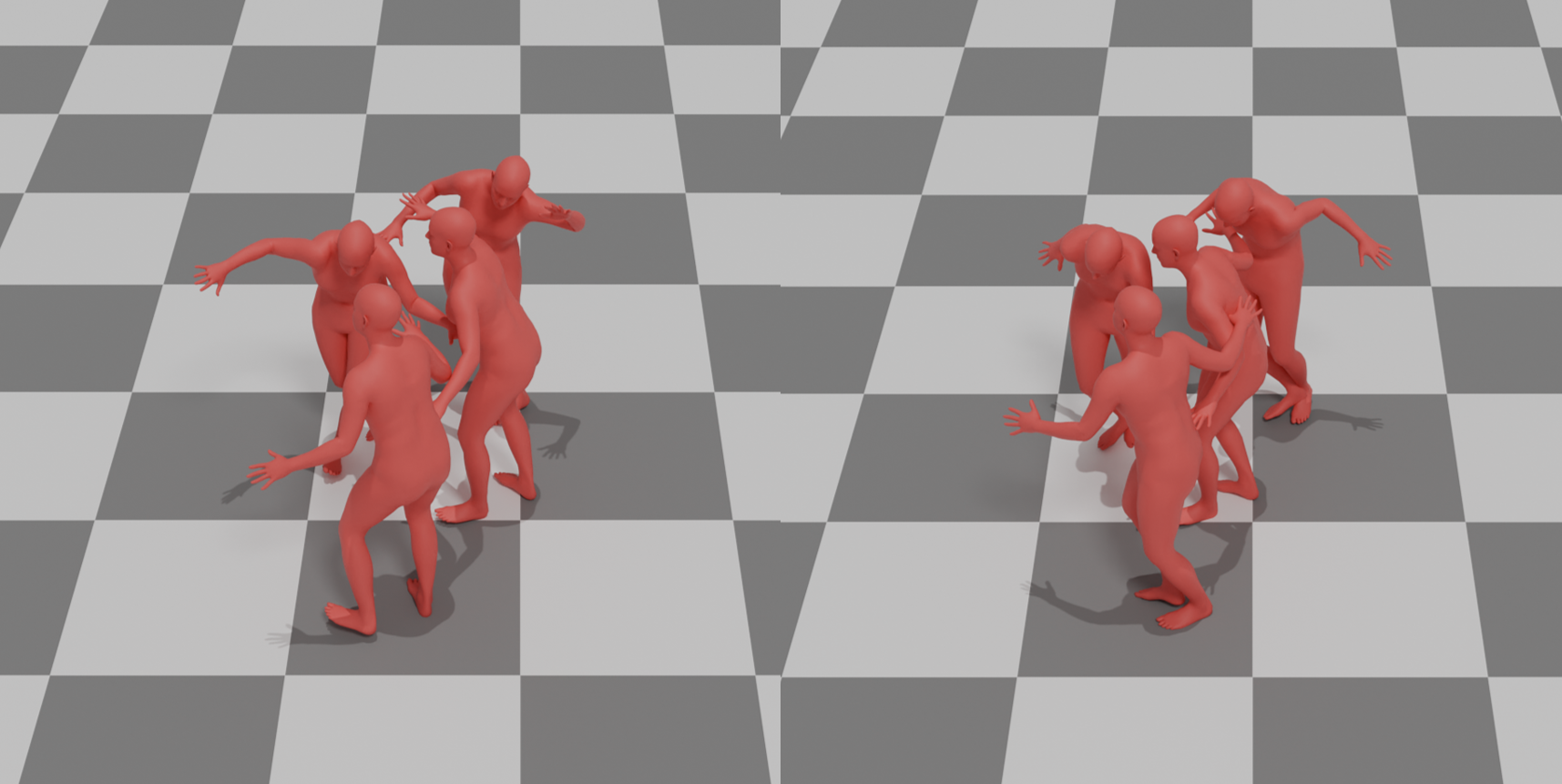}
    \caption{4-person dancing generated by ours}
    \label{fig:demo_b}
  \end{subfigure}
  \caption{Qualitative results from FreeMotion and our method. The prompt used in (a) and (b) is ``Three person attack each other with their punch''. The text used for generating (c) and (d) is ``They initiate a dance routine that involves swaying shoulders and arms''.}
  \label{fig:demo}
\end{figure*}

\begin{figure*}
  \setlength{\abovecaptionskip}{2pt}
  \centering
    \begin{subfigure}{0.9\linewidth}
    \includegraphics[width=\linewidth,trim={0 2.4cm 0 2.0cm},clip]{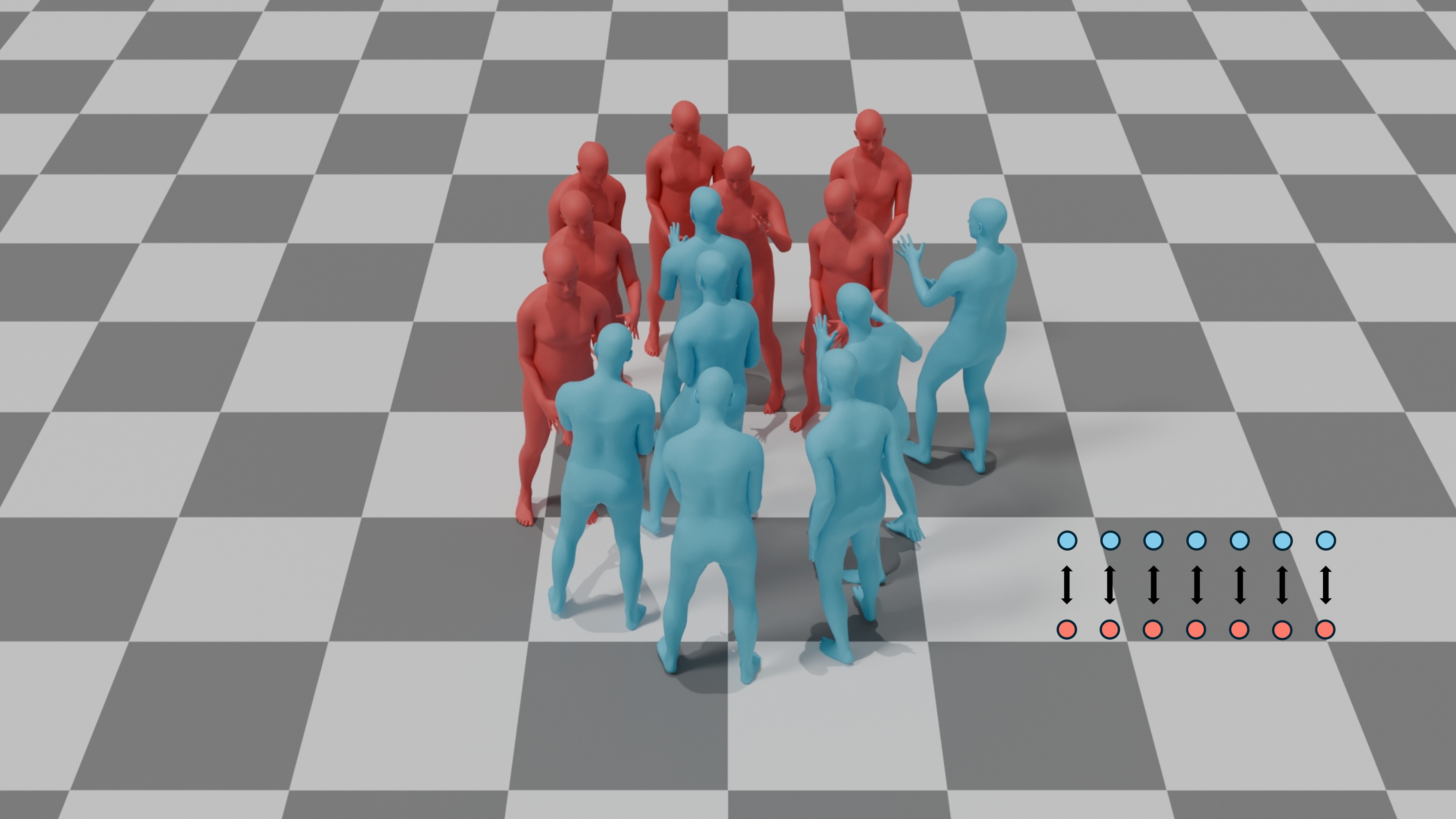}
    \caption{``People blame each other'' -- Two teams are negotiating. }
    \label{fig:supp1}
  \end{subfigure}
  \begin{subfigure}{0.45\linewidth}
    \includegraphics[width=\linewidth,trim={0 0cm 0 1.cm},clip]{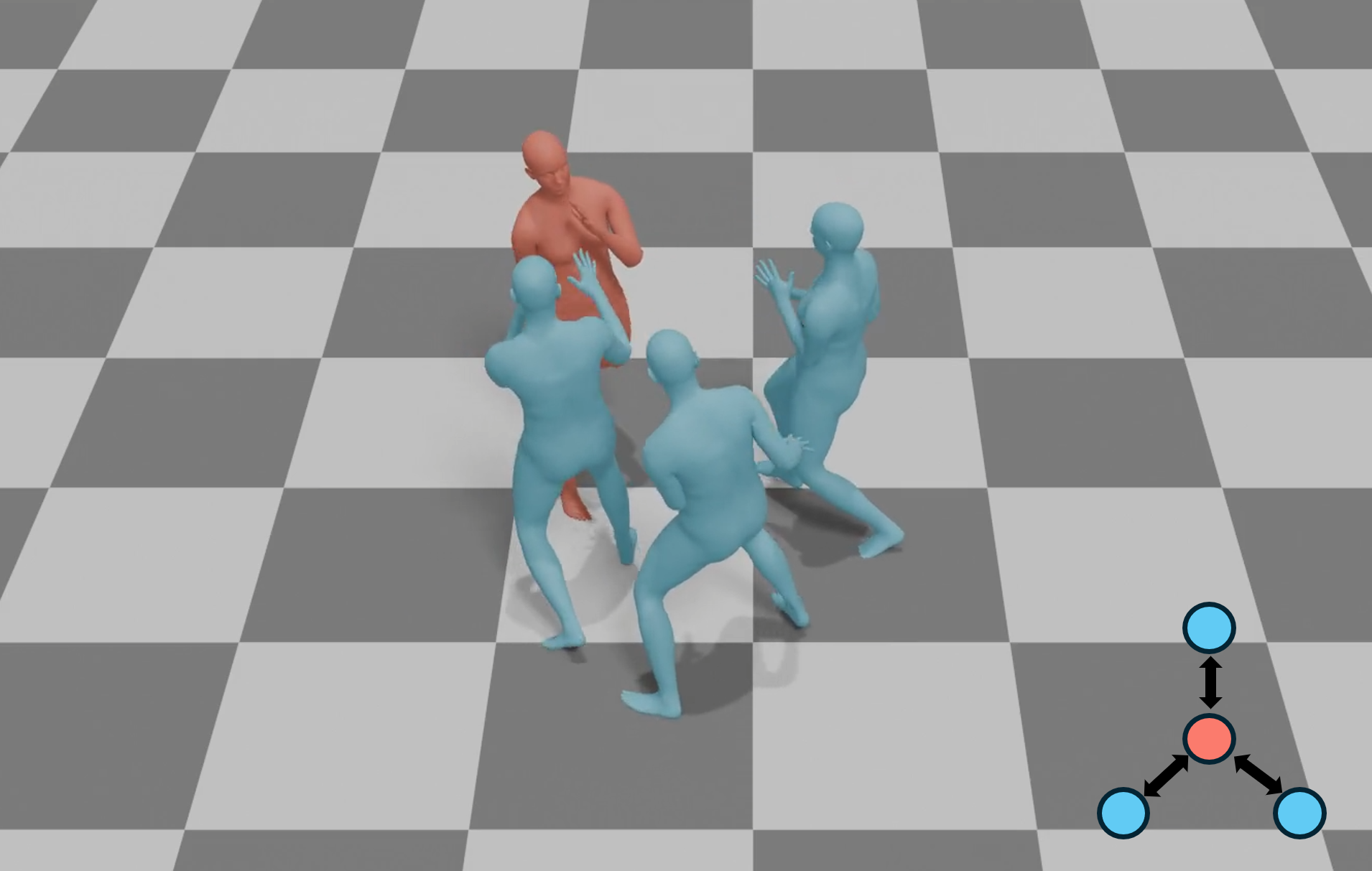}
    \caption{``People punching each other''-- the graph specifies that the three blue characters attack the red character}
    \label{fig:supp2}
  \end{subfigure}
    \begin{subfigure}{0.45\linewidth}
    \includegraphics[width=\linewidth,trim={0 0cm 0 1.cm},clip]{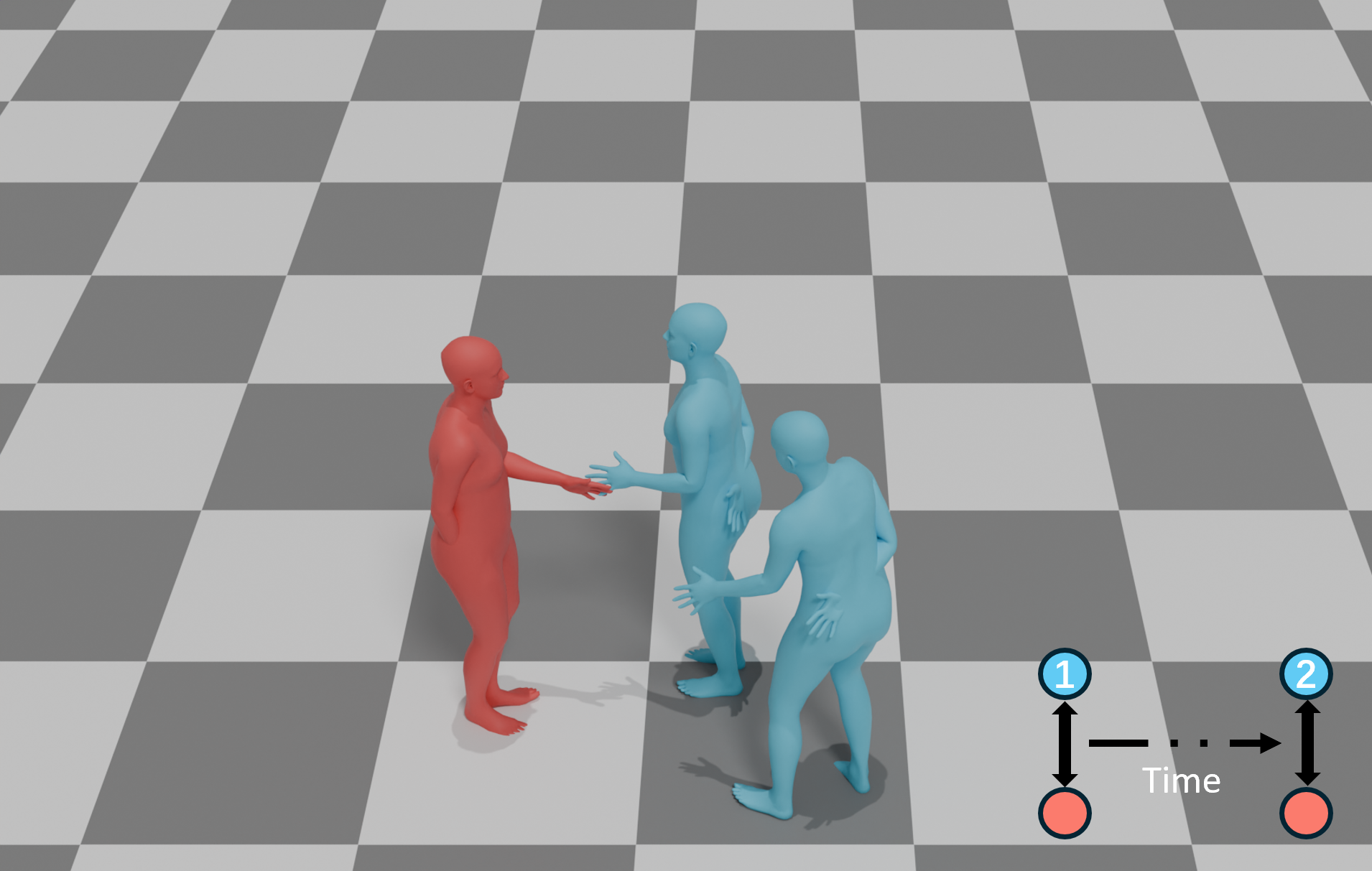}
    \caption{ ``People greet and handshake''-- time-varying sampling, the red character first handshakes with blue character \#1 then \#2. }
    \label{fig:supp3}
  \end{subfigure}
  \caption{More qualitative results on multi-person interactions generation}
  \label{fig:more_multi}
\end{figure*}

\begin{figure*}
  \setlength{\belowcaptionskip}{6pt}
  \setlength{\abovecaptionskip}{0pt}
  \centering
    \begin{subfigure}{0.9\linewidth}
    \includegraphics[width=\linewidth]{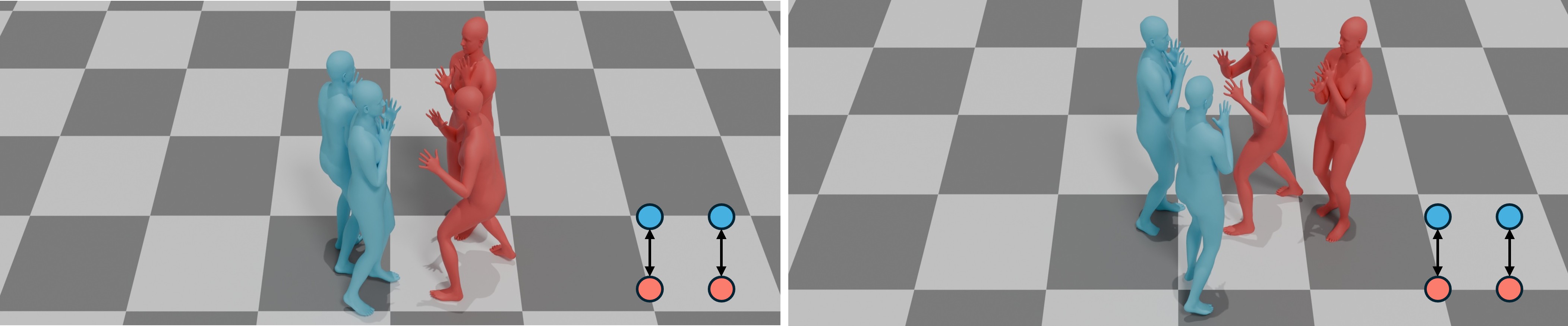}
    \caption{``People punching each other''-- generated by using MDM trained on InterHuman as our backbone. }
    \label{fig:supp4}
  \end{subfigure}
    \begin{subfigure}{0.9\linewidth}
    \includegraphics[width=\linewidth]{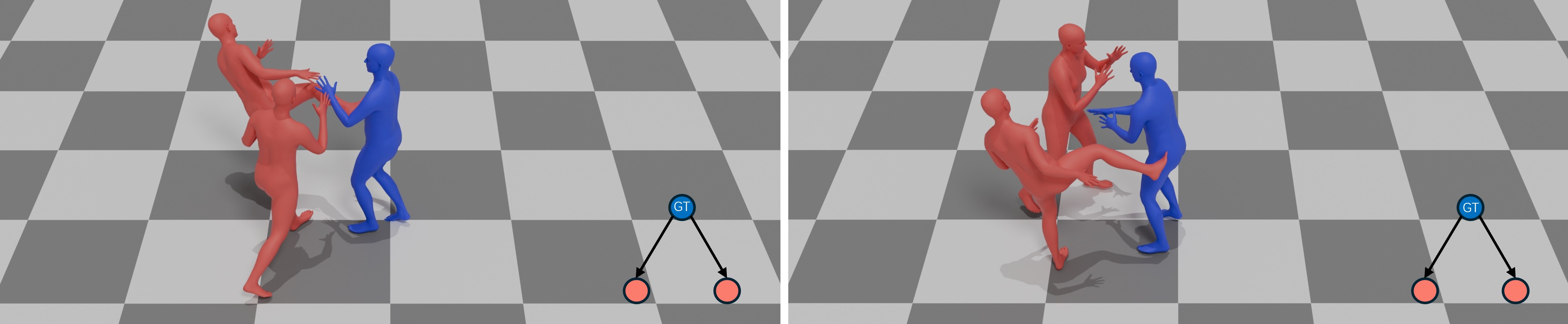}
    \caption{Multiple reaction generation using FreeMotion as our backbone-- conditions of red characters are the clean motion of the blue character, who is being kicked. }
    \label{fig:supp5}
  \end{subfigure}
    \begin{subfigure}{0.9\linewidth}
    \includegraphics[width=\linewidth]{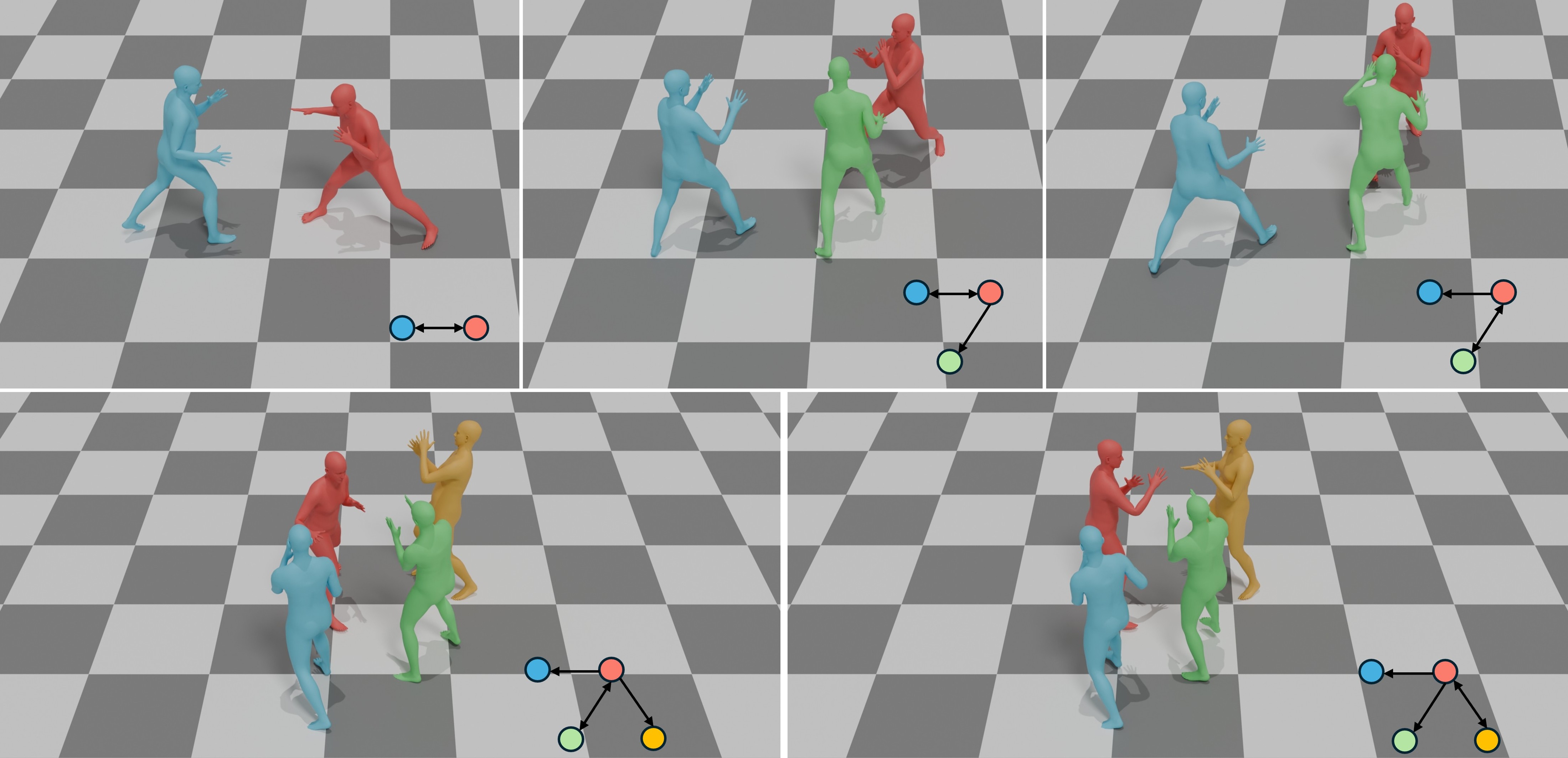}
    \caption{``People attack each other with punch''-- time-varying sampling, the red character changes target between more enemies. }
    \label{fig:supp6}
  \end{subfigure}
  \caption{More qualitative results on multi-person interactions generation}
  \label{fig:more_multi_2}
\end{figure*}

\begin{figure*}
  \setlength{\abovecaptionskip}{2pt}
  \centering
  \begin{subfigure}{0.49\linewidth}
    \includegraphics[width=\linewidth,trim={0 1.cm 0 1.cm},clip]{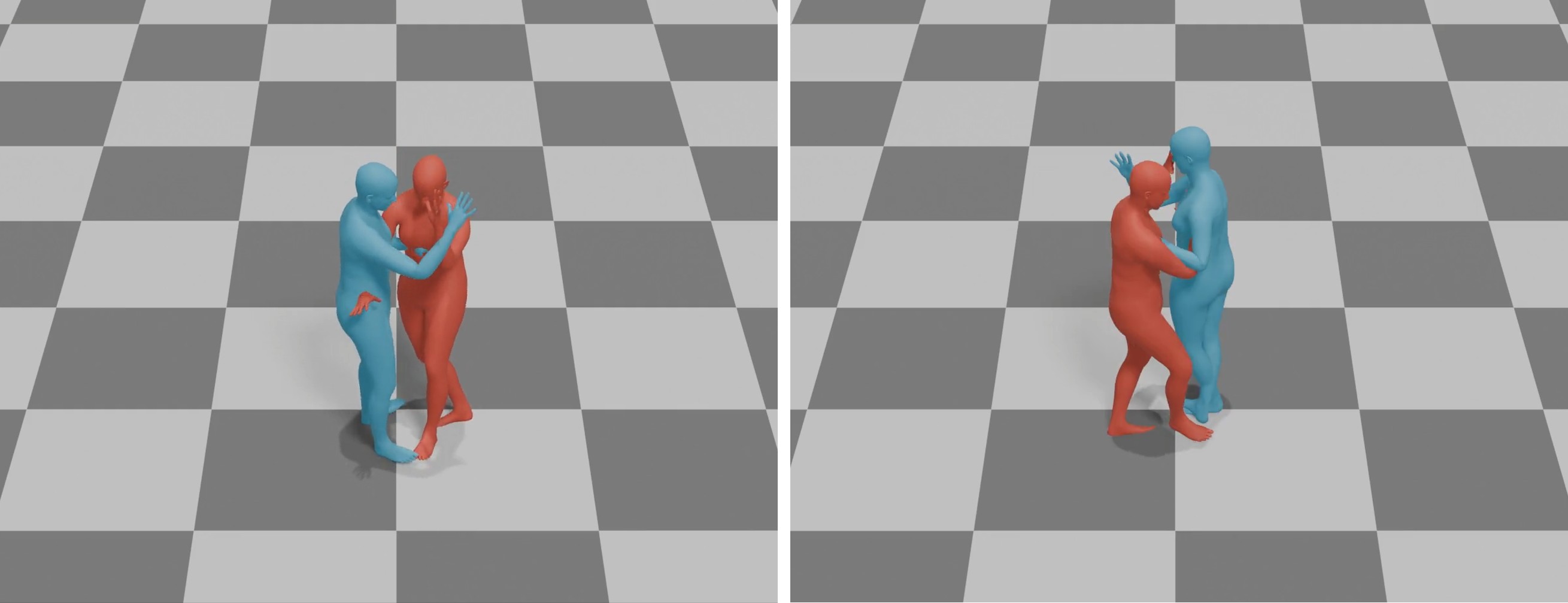}
    \caption{Results from InterGen}
    \label{fig:togli1}
  \end{subfigure}
\hfill
\begin{subfigure}{0.49\linewidth}
    \includegraphics[width=\linewidth,trim={0 1.cm 0 1.cm},clip]{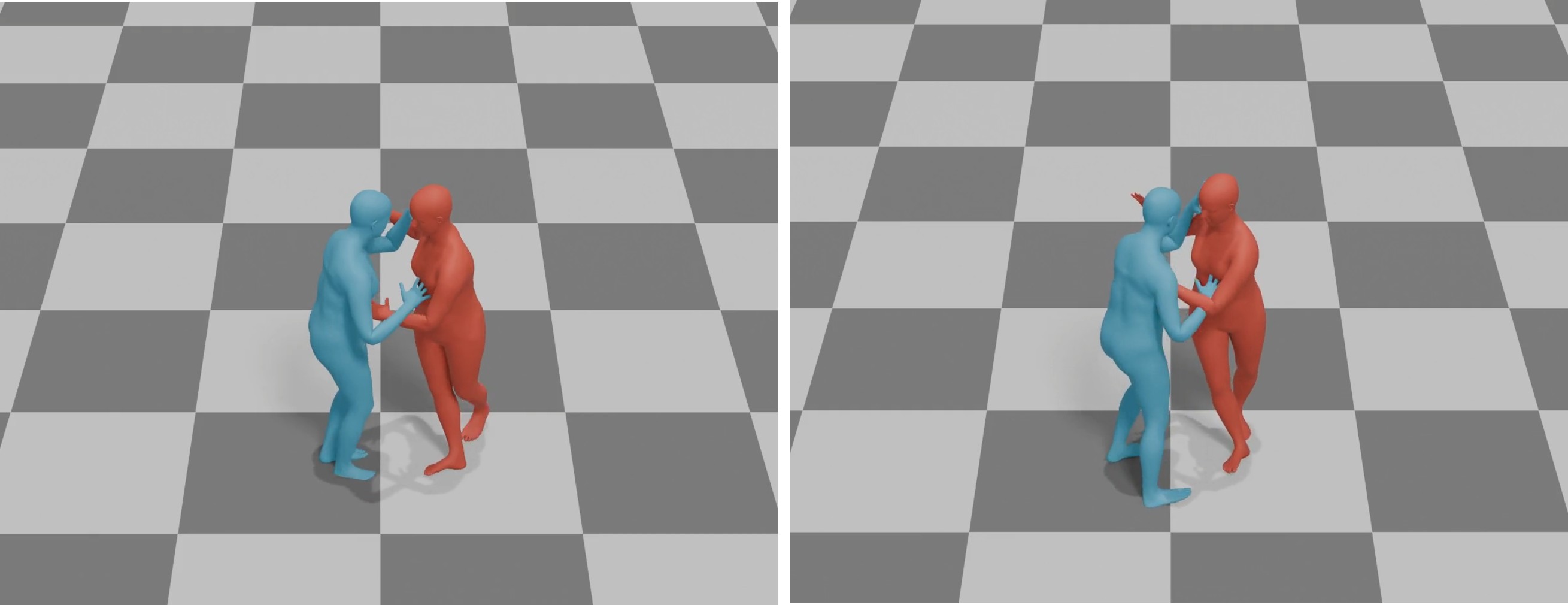}
    \caption{Results from ours}
    \label{fig:togli2}
  \end{subfigure}
  \caption{Qualitative results of 2-person interactions. The prompt used is ``one person embraces the other person and dances with joy.''.}
  \label{fig:togli}
\end{figure*}

%% file: sec/X_suppl.tex
\section{Implementation details}
For Sampling Optimization in DDPM-1000, we apply Proxemics loss when the timestep is smaller than 700 and GLI loss when the timestep is lower than 100. We optimize all
50 timesteps for DDIM-50. We use DDIM on two-person interaction
sampling and DDPM for multi-person.

\section{Evaluation Metrics}
We use four new metrics to evaluate the penetration magnitude in two-person and multi-person interactions. As all metrics are to measure the penetration of two-person, we take all combinations in a multi-person interaction for evaluation. To evaluate the skeleton level penetration, \textit{PeneBone} is introduced. We create the mesh of each 'bone' in the form of a box with a radius of 2cm. Then, for each bone, we detect if there is any contact with others' bones. We sum over all the collision depths detected in this way for each sample. For \textit{Pair-FID}, we evaluate the distribution of each pair in the multi-person interactions to several repetitive ground truth two-person interactions. To more accurately evaluate the collisions in the generated motions during practical use, we fitted the generated results to the SMPL model \cite{loper2023smpl}. However, it is important to note that since InterGen is trained on joint-based motion, this could lead to some issues. Similar to PeneBone, we counted the number of mesh collision points for each pair. Additionally, we calculated the proportion of frames with collisions relative to the total number of frames as a more meaningful metric. These two metrics based on smpl mesh are denoted as \textit{Contact} and \textit{CFrame}.

\section{ Close Two-person Interaction Subset}
Since the InterHuman dataset \cite{liang2024intergen} is not specifically focused on close interactions, it contains many completely separated actions, such as waving or playing games, which can impact the evaluation of penetration. Therefore, we selected a subset of close interactions from the test set that involve contact. We first filtered out motions likely to involve interactions based on the text prompts. Then, we selected all motions with more than 200 frames and chose the top 120 motions with the highest PeneBone values.

\section{Discretisation and Analytical Solution for GLI Calculation} \label{sec:dGLI}
As stated in Section 4.2.1, the GLI of two curves can be computed using Eq. 5. However, it is computationally costly to compute the GLI for long curves since there is a double integral in the equation. On the other hand, an analytical solution~\cite{Levitt:JMB1983} can be used when chained segments approximate the curves. Here, we first discretise the curves by representing each character using two sets of chains to capture the limb-level and body-level interactions as explained in Section 3.1.2 in the paper. 

Given two chains, $S1$ and $S2$, with $m$ and $n$ line segments on the chains, respectively. The GLI can be calculated by
\begin{eqnarray*}
GLI(S_1, S_2) = \sum_{i=1}^m \sum_{j=1}^n T_{i,j}
\end{eqnarray*}
where $T_{i,j}$ is the writhe of segment $i$ and segment $j$. 

The analytical solution~\cite{Levitt:JMB1983} can be used for calculating $T_{i,j}$. Specifically, given line segments $i$ (with end points $a$ and $b$) and $j$ (with endpoints $c$ and $d$), we can define 6 vectors as $v_{ab}$ ($a\rightarrow b$), $v_{ac}$ ($a\rightarrow c$), $v_{ad}$ ($a\rightarrow d$), $v_{bc}$ ($b\rightarrow c$), $v_{bd}$ ($b\rightarrow d$), $v_{cd}$ ($c\rightarrow d$). The vectors will be used for calculating the normal vectors of the tetrahedron created using those 4 endpoints:
\begin{eqnarray*}
n_{a} = \frac{v_{ac} \times v_{ad}}{\left|v_{ac} \times v_{ad}\right|},
n_{b} = \frac{v_{ad} \times v_{bd}}{\left|v_{ad} \times v_{bd}\right|},\\
n_{c} = \frac{v_{bd} \times v_{bc}}{\left|v_{bd} \times v_{bc}\right|},
n_{d} = \frac{v_{bc} \times v_{ac}}{\left|v_{bc} \times v_{ac}\right|}. 
\end{eqnarray*}
Finally, ${T_{i,j}}$ is calculated by     
\begin{align*}
T_{i,j} = \arcsin(n_an_b) + \arcsin(n_bn_c) + \arcsin(n_cn_d) \\+ \arcsin(n_dn_a).
\end{align*}



\section{Video Results}
Readers are referred to the video demo for more qualitative results discussed in the paper.